\RequirePackage{ifpdf}
\ifpdf
    \documentclass[pdftex,11pt,a4paper,oneside,final]{article}
\else
    \documentclass[dvips,11pt,a4paper,oneside,final]{article}
\fi

\newcommand{\myqed}{}

\usepackage[latin1]{inputenc}
\usepackage{a4wide}
\usepackage[pdftitle   = {From Strong Amalgamability to Modularity of
  Quantifier-Free Interpolation},
            pdfauthor  = {{Roberto ruttomesso - Silvio Ghilardi - Silvio Ranise }},
            pdfsubject = {{Rapporto Interno DSI 337-12}}]{}

\usepackage{amsmath,amsthm,amssymb}

\usepackage{ifthen}
\usepackage{graphicx}
\usepackage{fancybox}
\usepackage{boxedminipage}
\usepackage{enumerate}
\usepackage{alltt}
\usepackage{proof}
\usepackage{rotating,graphics}

\theoremstyle{plain}
\newtheorem{theorem}{Theorem}[section]
\newtheorem{lemma}[theorem]{Lemma}
\newtheorem{corollary}[theorem]{Corollary}

\newtheorem{proposition}[theorem]{Proposition}

\theoremstyle{definition}
\newtheorem{definition}[theorem]{Definition}

\theoremstyle{remark}

\usepackage{amsmath}
\usepackage{amstext}
\usepackage{amssymb}
\usepackage{amsfonts}
\usepackage{xspace}
\usepackage{graphicx}
\usepackage{url}
\usepackage{mdwlist}
\usepackage{courier}
\usepackage{lscape}
\usepackage{comment}
\usepackage{wrapfig}
\usepackage{multirow}
\usepackage{bussproofs}
\usepackage{pdftricks}
\begin{psinputs}
\usepackage{pstricks}
\usepackage{color}
\usepackage{pstcol}
\usepackage{pst-plot}
\usepackage{pst-tree}
\usepackage{pst-eps}
\usepackage{multido}
\usepackage{pst-node}
\usepackage{pst-eps}
\end{psinputs}
\usepackage{cite}
\makeatletter
\newif\if@restonecol
\makeatother

\usepackage[ruled,vlined]{algorithm2e}
\usepackage{footmisc}
\usepackage{balance}
\usepackage{enumerate}
\usepackage[normalem]{ulem}
\usepackage[a4paper]{hyperref}

\newcommand{\diff}{\ensuremath{\mathtt{diff}}\xspace}
\newcommand{\ARRAY}{\ensuremath{\mathtt{ARRAY}}\xspace}
\newcommand{\INDEX}{\ensuremath{\mathtt{INDEX}}\xspace}
\newcommand{\ELEM}{\ensuremath{\mathtt{ELEM}}\xspace}
\newcommand{\formulae}{formul\ae\xspace}

\newcommand{\PrA}{\ensuremath{\mathcal{PRA}}\xspace}

\newcommand{\EUF}{\ensuremath{\mathcal{EUF}}\xspace}
\newcommand{\LIA}{\ensuremath{\mathcal{LIA}}\xspace}
\newcommand{\IDL}{\ensuremath{\mathcal{IDL}}\xspace}
\newcommand{\AX}{\ensuremath{\mathcal{AX}}\xspace}
\newcommand{\UTVPI}{\ensuremath{\mathcal{UTVPI}}\xspace}

\newcommand{\AXEXT}{\ensuremath{\mathcal{AX}_{{\rm ext}}}\xspace}
\newcommand{\AXDIFF}{\ensuremath{\AX_{\diff}}\xspace}

\newcommand{\imp}{\Rightarrow}

\newcommand{\dpll}[1]{{\sc DPLL}\xspace}

\newcommand{\sep}{\ensuremath{\ | \ }}

\newcommand{\COMMENT}[1]{}

\newcommand{\alocal}{\ensuremath{A}-local\xspace}
\newcommand{\blocal}{\ensuremath{B}-local\xspace}
\newcommand{\astrict}{\ensuremath{A}-strict\xspace}
\newcommand{\bstrict}{\ensuremath{B}-strict\xspace}
\newcommand{\abcommon}{\ensuremath{AB}-common\xspace}
\newcommand{\abpure}{\ensuremath{AB}-pure\xspace}
\newcommand{\abmixed}{\ensuremath{AB}-mixed\xspace}

\newcommand{\ua}{\ensuremath{\underline a}}
\newcommand{\ub}{\ensuremath{\underline b}}
\newcommand{\uc}{\ensuremath{\underline c}}

\newcommand{\um}{\ensuremath{\underline m}}
\newcommand{\un}{\ensuremath{\underline n}}
\newcommand{\ut}{\ensuremath{\underline t}}
\newcommand{\uv}{\ensuremath{\underline v}}

\newcommand{\ux}{\ensuremath{\underline x}}
\newcommand{\uy}{\ensuremath{\underline y}}
\newcommand{\uz}{\ensuremath{\underline z}}

\newcommand{\cA}{\ensuremath \mathcal A}

\newcommand{\cM}{\ensuremath \mathcal M}
\newcommand{\cN}{\ensuremath \mathcal N}

\newcommand{\uguale}{\ensuremath{=}\xspace}
\newcommand{\coincide}{\ensuremath{\equiv}\xspace}

\newcommand{\mydiv}[2]{\ensuremath{#1~\mathit{div}[#2]}}
\newcommand{\myrem}[2]{\ensuremath{#1~\mathit{rem}[#2]}}

\makeindex
\begin{document}

\title{From Strong Amalgamability to Modularity of
  Quantifier-Free Interpolation }
\author{Roberto Bruttomesso$^1$ and Silvio Ghilardi$^1$ and Silvio Ranise$^{2}$ 
\\ $^1$Dipartimento di Scienze dell'Informazione, Universit\`a degli Studi di Milano (Italy)
\\ $^1$FBK (Fondazione Bruno Kessler), Trento, (Italy)}
%

\date{\today}

\thispagestyle{empty}
\maketitle

\begin{abstract}
The use of interpolants in verification is gaining more and more
  importance.  Since theories used in applications are usually
  obtained as (disjoint) combinations of simpler theories, it is
  important to modularly re-use interpolation algorithms for the
  component theories.  We show that a sufficient
  and necessary condition to do this for quantifier-free interpolation is
  that the component theories have the `strong (sub-)amalgamation'
  property.  Then, we 
  provide an equivalent syntactic characterization, identify 
  a sufficient 
  condition, and 
  design a combined quantifier-free
  interpolation algorithm capable of handling both convex and non-convex theories, that subsumes and extends most existing work
  on combined interpolation.
\end{abstract}

\newpage
\tableofcontents
\newpage

\section{Introduction}
\label{sec:introduction}

Algorithms for computing interpolants are more and more used in
verification, e.g., in the abstraction-refinement phase of software
model checking~\cite{HJM+04}.  Of particular importance in practice
are those algorithms capable of computing \emph{quantifier-free}
interpolants in presence of some background theory.
Since theories commonly used in verification are obtained as
combinations of simpler theories, methods to modularly combine
available quantifier-free interpolation algorithms are desirable.
This paper studies the modularity of quantifier-free interpolation.

Our starting point is the well-known fact~\cite{amalgam} that
quantifier-free interpolation (for universal theories) is equivalent
to the model-theoretic property of \emph{amalgamability}.
Intuitively, a theory has the amalgamation property if any two
structures $\cM_1, \cM_2$ in its class of models sharing a common
sub-model $\cM_0$ can be regarded as sub-structures of a larger model
$\cM$, 
called the amalgamated model.  Unfortunately, this property is
not sufficient to derive a modularity result for quantifier-free
interpolation.  As shown in this paper, a stronger notion is needed,
called \emph{strong amalgamability}~\cite{SAP}, that has been
thouroughly analyzed in universal algebra and category
theory~\cite{ringel,tholen}.  A theory has the strong amalgamation
property if in the amalgamated model $\cM$, elements from the supports
of $\cM_1, \cM_2$ not belonging to the support of $\cM_0$ \emph{cannot
  be identified}.  An example of an amalgamable but not strongly
amalgamable theory is the theory of fields: 
let $\cM_0$ be a real field and $\cM_1, \cM_2$ be two copies of the
complex numbers, the imaginary unit in
$\cM_1$ must be identified with the imaginary unit of 
$\cM_2$ (or with its opposite) in any amalgamating field $\cM$ since
the polynomial $x^2+1$ cannot have more than two roots (more examples
will be discussed below, many examples are also supplied in the
catalogue of~\cite{tholen}).  We show that \emph{strong amalgamability
  is precisely what is needed for 
the modularity of quantifier-free interpolation}, in the following
sense (here, for simplicity, we assume that theories are universal
although in the paper we generalize to arbitrary ones): 
(\emph{a}) if $T_1$ and $T_2$ are signature disjoint, both stably
infinite and strongly amalgamable, then $T_1\cup T_2$ is also strongly
amalgamable
and hence quantifier-free interpolating and (\emph{b}) a theory $T$ is
strongly amalgamable iff the disjoint union of $T$ with the theory
$\EUF$ of equality with uninterpreted symbols has quantifier-free
interpolation (Section~\ref{sec:strong_amalagamation}).  The first two
requirements of (\emph{a}) are those for the correctness of the
Nelson-Oppen method~\cite{NO79} whose importance for combined
satisfiability problems is well-known.
 
Since the proof of (\emph{a}) is non-constructive, the result does not
provide an algorithm to compute quantifier-free interpolants in
combinations of theories.  To overcome this problem, we 
reformulate
the
notion of \emph{equality interpolating} theory $T$ in terms of the
capability of computing some terms that are equal to the variables
occurring in disjunctions of equalities entailed (modulo $T$) by pairs
of quantifier-free formulae and show that equality interpolation is
  \emph{equivalent} to strong amalgamation
(Section~\ref{sec:strong_amalagamation_syntactically}).  To put
{equality interpolation} to productive work, we show that 
\emph{universal}
theories admitting elimination of quantifiers are equality
interpolating (Section~\ref{subsec:eq-interpol-at-work}).  This
implies that the theories of recursively defined data
structures~\cite{O1}, Integer Difference Logic,
Unit-Two-Variable-Per-Inequality, and Integer Linear Arithmetic with
division-by-$n$~\cite{brillout}
are all equality interpolating.  Our notion of equality interpolation
is a strict generalization of the one in~\cite{ym} so that all the
theories that are equality interpolating in the sense of~\cite{ym} are
also so according to our definition, e.g., the theory of LISP
structures~\cite{NO79} and Linear Arithmetic over the Reals
(Section~\ref{subsec:comparison}).  Finally, we describe a combination
algorithm for the generation of quantifier-free interpolants from
finite sets of quantifier-free formulae in unions of signature
disjoint, stably infinite, and equality interpolating theories
(Section~\ref{sec:algorithms}).  The algorithm uses as sub-modules the
interpolation algorithms of the component theories and is based on a
sequence of syntactic manipulations organized in groups of syntactic
transformations modelled after a non-deterministic version of the
Nelson-Oppen combination schema (see, e.g.,~\cite{TinHar-FROCOS-96}).
%
%
All the proofs are in Appendix~\ref{app:proofs}. The other Appendixes contain additional information on related topics, in particular
Appendix~\ref{app:beth} connects equality interpolation with Beth definability property, Appendix~\ref{app:general} investigates interpolation in presence of free
function symbols and Appendix~\ref{app:counterexample}
supplies a formal counterexample showing that the convex formulation of the equality interpolation property is insufficient to guarantee combined 
quantifier-free interpolation for non-convex theories.

\section{Formal Preliminaries}
\label{sec:background}

We assume the usual syntactic
and semantic
notions of first-order logic (see, e.g.,~\cite{enderton}).  The
equality symbol ``\uguale'' is included in all signatures considered
below.  For clarity, we shall use ``\coincide'' in the meta-theory to
express the syntactic identity between two symbols or two strings of
symbols.  Notations like $E(\ux)$ means that the expression (term,
literal, formula, etc.) $E$ contains free variables only from the
tuple $\ux$.  A `tuple of variables' is a list of variables without
repetitions and a `tuple of terms' is a list of terms (possibly with
repetitions).  Finally, whenever we use a notation like $E(\ux, \uy)$
we implicitly assume not only that both the $\ux$ and the $\uy$ are
pairwise distinct, but also that $\ux$ and $\uy$ are disjoint.  A
formula is \emph{universal} (\emph{existential}) iff it is obtained
from a quantifier-free formula by prefixing it with a string of
universal (existential, resp.)  quantifiers.

\noindent {\textbf{Theories, elimination of quantifiers, and
    interpolation}.}  A \emph{theory} $T$ is a pair $({\Sigma},
Ax_T)$, where $\Sigma$ is a signature and $Ax_T$ is a set of
$\Sigma$-sentences, called the \emph{axioms} of $T$ (we shall
sometimes write directly $T$ for $Ax_T$).  The \emph{models} of $T$
are those $\Sigma$-structures in which all the sentences in $Ax_T$ are
true.  
A $\Sigma$-formula $\phi$ is \emph{$T$-satisfiable} if there exists a
model $\cM$ of $T$ such that $\phi$ is true in $\cM$ under a suitable
assignment $\mathtt a$ to the free variables of $\phi$ (in symbols,
$(\cM, \mathtt a) \models \phi$); it is \emph{$T$-valid} (in symbols,
$T\vdash \varphi$) if its negation is $T$-unsatisfiable or,
equivalently, $\varphi$ is provable from the axioms of $T$ in a
complete calculus for first-order logic.  
A theory $T=(\Sigma, Ax_T)$ is \emph{universal} iff there is a theory
$T'=(\Sigma, Ax_{T'})$ such that all sentences in $Ax_{T'}$ are
universal and the sets of $T$-valid and $T'$-valid sentences coincide.
A formula $\varphi_1$ \emph{$T$-entails} a formula $\varphi_2$ if
$\varphi_1 \to \varphi_2$ is \emph{$T$-valid} (in symbols,
$\varphi_1\vdash_T \varphi_2$ or simply $\varphi_1 \vdash \varphi_2$
when $T$ is clear from the context).  The \emph{satisfiability modulo
  the theory $T$} ($SMT(T)$) \emph{problem} amounts to establishing
the $T$-satisfiability of quantifier-free $\Sigma$-formulae.

A theory $T$ admits \emph{quantifier-elimination} iff for every
formula $\phi(\ux)$ there is a quantifier-free formula
$\phi'(\ux)$ such that $T\vdash \phi \leftrightarrow \phi'$.  A theory
$T$ \emph{admits quantifier-free interpolation} (or, equivalently,
\emph{has quantifier-free interpolants}) iff for every pair of
quantifier-free formulae $\phi, \psi$ such that $\psi\wedge \phi$ is
$T$-unsatisfiable, there exists a quantifier-free formula $\theta$,
called an \emph{interpolant}, such that: (i) $\psi$ $T$-entails
$\theta$, (ii) $\theta \wedge \phi$ is $T$-unsatisfiable, and (iii)
only the variables occurring in both $\psi$ and $\phi$ occur in
$\theta$.  
A theory admitting quantifier
elimination also admits quantifier-free interpolantion; the vice versa
does not hold. A more general notion of quantifier-free interpolation property, involving free function symbols, 
is discussed in Appendix~\ref{app:general}.

\noindent {\textbf{Embeddings, sub-structures, and combinations of
    theories}}.  The support of a structure $\mathcal{M}$ is denoted
with $|\mathcal{M}|$.  An embedding is a homomorphism that preserves
and reflects relations and operations (see, e.g.,~\cite{CK}).
Formally, a {\it $\Sigma$-embedding} (or, simply, an embedding)
between two $\Sigma$-structu\-res $\cM$ and $\cN$ is any mapping $\mu:
|\cM| \longrightarrow |\cN|$ satisfying the following three
conditions: (a) it is a
injective function; (b) it is an algebraic homomorphism, that is for
every $n$-ary function symbol $f$ and for every $a_1, \dots, a_n\in
|\mathcal{M}|$, we have $f^{\cN}(\mu(a_1), \dots, \mu(a_n))=
\mu(f^{\cM}(a_1, \dots, a_n))$; (c) it preserves and reflects
interpreted predicates, i.e.\ for every $n$-ary predicate symbol $P$,
we have $(a_1, \dots, a_n)\in P^{\cM}$ iff $(\mu(a_1), \dots,
\mu(a_n))\in P^{\cN}$.  If $|\mathcal{M}|\subseteq |\mathcal{N}|$ and
the embedding $\mu: \cM \longrightarrow \cN$ is just the identity
inclusion $|\mathcal{M}|\subseteq |\mathcal{N}|$, we say that $\cM$ is
a {\it substructure} of $\cN$ or that $\cN$ is a {\it superstructure}
of $\cM$.
%
%
As it is well-known, the truth of a universal (resp. existential)
sentence is preserved through substructures (resp. superstructures).

A theory $T$ is \emph{stably infinite} iff every $T$-satisfiable
quantifier-free formula (from the signature of $T$) is satisfiable in
an infinite model of $T$.  By compactness, it is possible to show that
$T$ is {stably infinite} iff every model of $T$ embeds into an
infinite one (see~\cite{ghilardi-jar}).  A theory $T$ is \emph{convex} iff
for every conjunction of literals $\delta$, if $\delta\vdash_T
\bigvee_{i=1}^n x_i=y_i$ then $\delta\vdash_T x_i=y_i$ holds for some
$i\in \{1, ..., n\}$.

Let $T_i$ be a stably-infinite theory over the signature $\Sigma_i$
such that the $SMT(T_i)$ problem is decidable for $i=1,2$ and
$\Sigma_1$ and $\Sigma_2$ are disjoint (i.e.\ the only shared symbol
is equality).  Under these assumptions, the Nelson-Oppen combination
method~\cite{NO79} tells us that the SMT problem for the combination
$T_1\cup T_2$ of the theories $T_1$ and $T_2$ (i.e.\ the union of
their axioms)
 is
decidable. 

\section{Strong amalgamation and quantifier-free interpolation}
\label{sec:strong_amalagamation}

We first generalize the notions of amalgamability and strong
amalgamability to arbitrary theories.
\begin{definition}
  \label{def:sub-amalgamation-and-strong-sub-amalgamation}
  A theory $T$ has the \emph{sub-amalgamation property} iff whenever
  we are given models $\cM_1$ and $\cM_2$ of $T$ and a common
  substructure $\cA$ of them, there exists a further model $\cM$ of
  $T$ endowed with embeddings $\mu_1:\cM_1 \longrightarrow \cM$ and
  $\mu_2:\cM_2 \longrightarrow \cM$ whose restrictions to $|\cA|$
  coincide.\footnote{For the results of this paper to be correct, the
    notion of structure (and of course that of substructure) should
    encompass the case of structures with empty domains.
    Readers feeling unconfortable with empty domains can assume that
    signatures always contain an individual constant.  }

  A theory $T$ has the \emph{strong sub-amalgamation property} if the
  embeddings $\mu_1, \mu_2$ satisfy the following additional
  condition: if for some $m_1, m_2$ we have $\mu_1(m_1)=\mu_2(m_2)$,
  then there exists an element $a$ in $|\cA|$ such that $m_1=a=m_2$.
\end{definition}
If the theory $T$ is universal, any substructure of a model of $T$ is
also a model of $T$ and we can assume that the substructure $\cA$ in
the definition above is also a model of $T$.  
In this sense,
Definition~\ref{def:sub-amalgamation-and-strong-sub-amalgamation}
introduces generalizations of the standard notions of amalgamability
and strong amalgamability for universal theories (see,
e.g.,~\cite{tholen} for a survey). 
The result of~\cite{amalgam} relating universal theories and
quantifier-free interpolation can be easily extended.  
\begin{theorem}
  \label{thm:interpolation-amalgamation} 
  A theory $T$ has the sub-amalgamation property iff it admits
  quan\-ti\-fier-free interpolants.
\end{theorem}
A theory admitting quantifier
elimination has the sub-amalgamation property: 
this follows, e.g., from
  Theorem~\ref{thm:interpolation-amalgamation} 
above. On the other hand,
quantifier elimination is not sufficient to guarantee the strong
sub-amalgamation property.  In fact, from
Theorem~\ref{prop:strong_amalgamation_needed} below and the
counterexample given in~\cite{BKR+2010}, it follows that Presburger
arithmetic does not have the strong sub-amalgamation property.
However, in Section~\ref{sec:strong_amalagamation_syntactically}, we
shall see that it is sufficient to enrich the signature of Presburger
Arithmetic with (integer) division-by-$n$ (for every $n\geq 1$) to
have strong amalgamability.

\noindent \textbf{Examples}.  For any signature $\Sigma$, let
$\EUF(\Sigma)$ be the pure equality theory over $\Sigma$.  It is easy
to see that $\EUF(\Sigma)$ is universal and has the strong
amalgamation property by building a model $\cM$ of $\EUF(\Sigma)$ from
two models $\cM_1$ and $\cM_2$ sharing a substructure $\cM_0$ as
follows. 
Without loss of generality, assume that $\vert \cM_0\vert =\vert \cM_1\vert \cap \vert \cM_2\vert$;
let $|\cM|$ be $\vert\cM_1\vert \cup \vert\cM_2\vert$ and
arbitrarily extend the interpretation of the function and predicate
symbols to make them total on $|\cM|$.

Let us now consider two variants \AXEXT 
and \AXDIFF of the theory of
arrays considered in~\cite{RTA}.  The signatures of \AXEXT and \AXDIFF
contain the sort symbols $\ARRAY, \ELEM$, and $\INDEX$, and the
function symbols $rd:\ARRAY \times \INDEX \longrightarrow \ELEM$ and
$wr:\ARRAY \times \INDEX \times \ELEM\longrightarrow \ARRAY$.  The
signature of \AXDIFF also contains the function symbol $\diff:
\ARRAY\times \ARRAY \longrightarrow \INDEX$.  The set \AXEXT of axioms
contains the following three sentences:
\begin{eqnarray*}
 \forall y, i, j, e. ~ i \not\uguale j \imp rd(wr(y,i,e),j)\uguale rd(y,j), 
  &~~~~~ &
 \forall y, i, e.~rd(wr(y,i,e),i) \uguale e, 
  \\
  \forall x, y. ~ 
  x \not\uguale y \imp (\exists i.\ rd(x,i)\not\uguale rd(y,i)) 
& &
\end{eqnarray*}
whereas 
the set  of axioms for \AXDIFF 
is obtained from that of \AXEXT by replacing the third axiom with its
 Skolemization:
\begin{eqnarray*}
  \forall x, y. & & 
  x \not\uguale y \imp rd(x,\diff(x,y))\not\uguale rd(y,\diff(x,y)) ~.~~~~~~~~~~~~
\end{eqnarray*} 
In~\cite{rta-extended} (the extended version of~\cite{RTA}), it is
shown that \AXDIFF has the strong sub-amalgamation property while
\AXEXT does not. However \AXEXT (which is \emph{not} universal) enjoys
the 
following 
property (this is the standard notion of amalgamability from the literatrure):
given two models $\cM_1$ and $\cM_2$ of \AXEXT
sharing a substructure $\cM_0$ \emph{which is also a model of} \AXEXT,
there is a model $\cM$ of \AXEXT endowed with embeddings from $\cM_1,
\cM_2$ agreeing on the support of $\cM_0$.


The application of Theorem~\ref{thm:interpolation-amalgamation} to
$\EUF(\Sigma)$, \AXDIFF, and \AXEXT allows us to derive in a uniform
way results about quantifier-free interpolation that are available in
the literature: that $\EUF(\Sigma)$ (see, e.g.,~\cite{McM04b,KFG+09})
and \AXDIFF~\cite{RTA} admit quantifier-free interpolants, and that
\AXEXT does not~\cite{KMZ06}.

\subsection{Modularity of quantifier-free interpolation}
\label{subsec:modularity-qf-interpol}
Given the importance of combining theories in SMT solving, the next
step is to establish whether sub-amalgamation is a modular property.
Unfortunately, this is not the case since the combination of two
theories having quantifier-free interpolation may not have
quantifier-free interpolation.  For example, the union of the theory
$\EUF(\Sigma)$ 
and Presburger
arithmetic 
does not have quantifier-free
interpolation~\cite{BKR+2010}.  Fortunately, strong sub-amalgamation
is modular when combining stably infinite theories.
\begin{theorem}
 \label{thm:sub-amalgamation_transfer}
 Let $T_1$ and $T_2$ be two stably infinite theories over disjoint
 signatures $\Sigma_1$ and $\Sigma_2$.  If both $T_1$ and $T_2$ have
 the strong sub-amalgamation property, then so does $T_1\cup T_2$.
\end{theorem}
Theorems~\ref{thm:interpolation-amalgamation}
and~\ref{thm:sub-amalgamation_transfer} obviously imply that strong
sub-amalgamation is sufficient for the modularity of quantifier-free
interpolation for stable infinite theories.
\begin{corollary}
  \label{coro:sub-amalgmation-sufficient}
  Let $T_1$ and $T_2$ be two stably infinite theories over disjoint
  signatures $\Sigma_1$ and $\Sigma_2$.  If both $T_1$ and $T_2$ have
  the strong sub-amalgamation property, then $T_1\cup T_2$ admits
  quantifier-free interpolation.  
\end{corollary}
We can also show that strong sub-amalgamation is necessary as
explained by the following result.
\begin{theorem}
  \label{prop:strong_amalgamation_needed} 
  Let $T$ be a theory admitting quantifier-free interpolation and
  $\Sigma$ be a signature disjoint from the signature of $T$
  containing at least a unary predicate symbol.  Then, $T\cup
  \EUF(\Sigma)$ has quantifier-free interpolation iff $T$ has the
  strong sub-amalgamation property.
\end{theorem}
Although Corollary~\ref{coro:sub-amalgmation-sufficient} is already
useful to establish whether 
combinations of theories admit quantifier-free interpolants, proving
the strong sub-amalgamability property can be complex.  In the next
section, we study an alternative (``syntactic'') characterization of
strong sub-amalgamability that can be more easily applied to commonly
used theories.

\section{Equality interpolation and strong amalgamation}
\label{sec:strong_amalagamation_syntactically}

There is a tight relationship between the strong sub-amalgamation
property of a theory $T$ and the fact that disjunctions of equalities
among variables are entailed by $T$.
 To state this precisely,
we need to introduce some preliminary notions.  Given two finite
tuples $\ut\coincide t_1, \dots, t_n$ and $\uv\coincide v_1,\dots,
v_m$ of terms,
\begin{eqnarray*}
  \text{the notation }
  \ut \cap \uv \neq \emptyset 
   \text{ stands for the formula } 
  \bigvee_{i=1}^n \bigvee_{j=1}^m (t_i\uguale v_j) .
\end{eqnarray*} 
We use $\ut_1\ut_2$ to denote the juxtaposition of the two tuples
$\ut_1$ and $\ut_2$ of terms.  So, for example, $\ut_1\ut_2\cap \uv
\not= \emptyset$ is equivalent to $(\ut_1 \cap \uv \not= \emptyset)
\vee (\ut_2 \cap \uv \not= \emptyset)$.
\begin{definition}
  \label{def:eq-interpol}
  A theory $T$ is \emph{equality interpolating} iff it has the
  quantifier-free interpolation property and satisfies the following
  condition:
  \begin{itemize}
    \item
      for every quintuple $\ux, \uy_1,\uz_1, \uy_2, \uz_2$ of
      tuples of variables and pair of quantifier-free formulae
      $\delta_1(\ux, \uz_1,\uy_1)$ and $\delta_2(\ux,\uz_2, \uy_2)$
      such that
      \begin{equation}
        \label{eq:antecedent_var}
        \delta_1(\ux,\uz_1, \uy_1) \wedge 
        \delta_2(\ux,\uz_2, \uy_2) \vdash_T 
        \uy_1\cap \uy_2\neq \emptyset
      \end{equation}
      there exists a tuple $\uv(\ux)$ of terms such that
      \begin{equation}
        \label{eq:consequent_var}
        \delta_1(\ux, \uz_1,\uy_1)\wedge 
        \delta_2(\ux,\uz_2, \uy_2)\vdash_T 
        \uy_1\uy_2\cap \uv\neq \emptyset ~.
      \end{equation} 
  \end{itemize}
\end{definition}
We are now in the position to formally state the equivalence between
strong sub-amalgamation and equality interpolating property.
\begin{theorem}
\label{thm:strong_amalgamation_syntactic}
A theory
$T$ has the  strong sub-amalgamation property iff it is equality interpolating.
 %
\end{theorem}

\subsection{Equality interpolation at work}
\label{subsec:eq-interpol-at-work}


We now illustrate some interesting applications of
Theorem~\ref{thm:strong_amalgamation_syntactic} so that, by using
Corollary~\ref{coro:sub-amalgmation-sufficient}, we can establish when
combinations of theories admit quantifier-free interpolation.
To ease the application of
Theorem~\ref{thm:strong_amalgamation_syntactic}, we first study the
relationship between quantifier-elimination and equality interpolation
for universal theories. 
\begin{theorem}
  \label{thm:qe_strong_amalgamability}
  A universal theory admitting quantifier elimination is equality
  interpolating.
\end{theorem}
Interestingly, the proof of this theorem (see
Appendix~\ref{subapp:proofs-of-sec:strong_amalagamation_syntactically})
is constructive and shows how an available quantifier elimination
algorithm (for a universal theory) can be used \emph{to find the terms
  $\uv$ satisfying condition~\eqref{eq:consequent_var} of
  Definition~\ref{def:eq-interpol}}; this is key to the combined
interpolation algorithm presented in Section~\ref{sec:algorithms}
below.

\noindent \textbf{Examples}.  
The theory $\mathcal{RDS}$ of \emph{recursive data
  structures}~\cite{O1} consists of two unary function symbols $car$
and $cdr$ and a binary function symbol $cons$, and it is axiomatized
by the following infinite set of sentences:
\begin{align*}
 \forall x,y.car(cons(x,y)) = x,
 ~~~~\forall x,y.cdr(cons(x,y)) = y, \tag{CCC}
 \\
 \forall x,y.cons(car(x), cdr(x))=x,
 ~~~~~~~~~~~~~~~~~~~~~~\forall x.x\neq t(x)
\end{align*}
where $t$ is a term obtained by finitely many applications of $car$
and $cdr$ to the variable $x$ (e.g., $car(x)\neq x$, $cdr(cdr(x))\neq
x$, $cdr(car(x))\neq x$, and so on).  Clearly, $\mathcal{RDS}$ is
universal; the fact that it admits elimination of quantifiers is known
since an old work by Mal'cev~\cite{Ma}.


Following~\cite{enderton}, we define the theory $\mathcal{IDL}$ of
\emph{integer difference logic} to be the theory whose signature contains the constant symbol $0$,
the unary function symbols $succ$ and $pred$, and the binary predicate
symbol $<$, and which is axiomatized by adding to the irreflexivity, transitivity and linearity axioms for $<$
the following set of
sentences:
\begin{eqnarray*}
  \begin{array}{ll}
  \forall x.succ(pred(x)) = x, 
 & \quad 
  \forall x.pred(succ(x)) = x,\\
  \forall x,y.x<succ(y)\leftrightarrow (x< y\vee x=y),
&  \quad
  \forall x,y.pred(x)<y\leftrightarrow (x< y\vee x=y).  
  \end{array}
\end{eqnarray*}
$\mathcal{IDL}$ is universal and the fact that admits elimination of
quantifiers can be shown by adapting the procedure for a similar
theory of natural numbers with successor and ordering
in~\cite{enderton}.  The key observation is that the atoms of
$\mathcal{IDL}$ are equivalent to formulae of the form $i \bowtie
f^n(j)$ (for $n\in \mathbb{Z}$, $\bowtie\,\in \{=, <\}$) where $i,j$
are variables or the constant $0$, $f^0(j)$ is $j$, $f^{k}(j)$
abbreviates $succ(succ^{k-1}(j))$ when $k>0$ or $pred(pred^{k-1}(j))$
when $k<0$.  (Usually, $i \bowtie f^n(j)$ is written as $i-j\bowtie n$
or as $i\bowtie j+ n$ from which the name of ``integer difference
logic.'')  

The theory $\mathcal{LAI}$ of Linear Arithmetic over the Integers
contains the binary predicate symbol $<$, the constant symbols $0$ and
$1$, the unary function symbol $-$, the binary function symbol $+$ and
the unary function symbols $\mydiv{}{n}$ (integer division by $n$, for
$n>1$). As axioms, we take a set of sentences such that all true
sentences in the standard model of the integers can be derived.
This can be achieved for instance
by adding to the axioms for
totally ordered Abelian groups the following 
sentences
(below $\myrem{x}{n}$ abbreviates $x-n(\mydiv{x}{n})$,
moreover $kt$ denotes
the sum $t+\cdots +t$ having $k$ addends all equal to the term $t$ and $k$ stands for $k1$):
\begin{eqnarray*}
  \begin{array}{l}
 %
   0<1, \quad 
  \forall y.\neg(0<y\wedge y<1),
  ~~~\text{and}~~~
  \forall x.\myrem{x}{n}=0\vee \cdots\vee \myrem{x}{n}=n-1 \, .
  \end{array}
\end{eqnarray*}
%
%
$\mathcal{LAI}$ can be seen as a variant of Presburger Arithmetic
obtained by adding the functions \mydiv{}{n}
instead of the `congruence modulo $n$' relations (for $n=1, 2, 3,
...$), which are needed to have quantifier elimination (see,
e.g.,~\cite{enderton}).  For the application of
Theorem~\ref{thm:qe_strong_amalgamability}, the problem with adding
the `congruence modulo $n$' is that the resulting theory is not
universal.  Instead, $\mathcal{LAI}$ is universal and the fact that
admits elimination of quantifiers can be derived by adapting existing
quantifier-elimination procedures (e.g., the one in~\cite{enderton})
and observing that $x$ is congruent to $y$ modulo $n$ can be defined
as $\myrem{x}{n} = \myrem{y}{n}$ (more details can be found in
Appendix~\ref{subapp:lai}).
%


By Theorem~\ref{thm:qe_strong_amalgamability}, $\mathcal{RDS}$,
$\mathcal{IDL}$, and $\mathcal{LAI}$ are equality interpolating.  The
theory $\mathcal{UTVPI}$ of Unit-Two-Variable-Per-Inequality (see,
e.g.,~\cite{CGS08}) is also equality interpolating (for lack of space,
this is shown in Appendix~\ref{subapp:utvpi}).

\subsection{A comparison with the notion of equality interpolation in~\cite{ym}}
\label{subsec:comparison}
We now show that the notion of equality interpolating theories
proposed here reduces to that of~\cite{ym} when considering convex
theories.  
\begin{proposition}
  \label{prop:ym-eq-interpol}
  A convex theory $T$ having quantifier-free interpolation is equality
  interpolating iff for every pair $y_1, y_2$ of variables and for
  every pair of 
conjunctions of literals
$\delta_1(\ux,\uz_1,
  y_1), \delta_2(\ux,\uz_2,y_2)$ such that
  \begin{equation}
    \label{eq:ym_ant}
    \delta_1(\ux, \uz_1,y_1)\wedge
    \delta_2(\ux,\uz_2, y_2)\vdash_T
    y_1= y_2
  \end{equation}
  there exists a term $v(\ux)$ such that
  \begin{equation}
    \label{eq:ym_cons}
    \delta_1(\ux, \uz_1, y_1)\wedge
    \delta_2(\ux, \uz_2, y_2)\vdash_T 
    y_1= v  \wedge  y_2= v.
  \end{equation}
\end{proposition}
The implication $(\ref{eq:ym_ant})\Rightarrow (\ref{eq:ym_cons})$ is
exactly the definition of equality interpolation in~\cite{ym}.  
In the following, a convex 
quantifier-free interpolating
theory satisfying
$(\ref{eq:ym_ant})\Rightarrow (\ref{eq:ym_cons})$ will be called
\emph{YMc equality interpolating}.  By
Proposition~\ref{prop:ym-eq-interpol}, an YMc equality interpolating
(convex) theory is also equality interpolating according to
Definition~\ref{def:eq-interpol}.  For example, the theory
$\mathcal{LST}$ of list structures~\cite{NO79} contains the function
symbols of $\mathcal{RDS}$, a unary predicate symbol $atom$, and it is
axiomatized by the  axioms of $\mathcal{RDS}$ labelled
$\mathit{(CCC)}$ and the sentences: 
\begin{eqnarray*}
\begin{array}{l}
  \forall x,y.\neg atom(cons(x,y)), \quad
  \forall x.\neg atom(x)\to cons(car(x),cdr(x))=x\, .
\end{array}
\end{eqnarray*}
$\mathcal{LST}$ is a (universal) convex theory~\cite{NO79} that was
shown to be YMc equality interpolating in~\cite{ym}.
By Proposition~\ref{prop:ym-eq-interpol}, we conclude that
$\mathcal{LST}$ is equality interpolating in the sense of
Definition~\ref{def:eq-interpol}.
%
In~\cite{ym}, also Linear Arithmetic over the Reals ($\mathcal{LAR}$)
is shown to be YMc equality interpolating
%
(the convexity of $\mathcal{LAR}$ is well-known from linear algebra).
By Proposition~\ref{prop:ym-eq-interpol}, $\mathcal{LAR}$ is equality
interpolating in the sense of Definition~\ref{def:eq-interpol}.
%
The same result can be obtained from
Theorem~\ref{thm:qe_strong_amalgamability} above by identifying a set
of universal axioms for the theory and showing that they admit
quantifier elimination.  
For the axioms to be universal,
it is essential to include
\emph{multiplication by rational coefficients} in the signature of the
theory, i.e.\ the unary function symbols $q*\_$ for every $q\in
\mathbb{Q}$.  If this is not the case, the theory is not
sub-amalgamable and thus not equality interpolating:  to see this,
consider the embedding of the substructure $\mathbb{Z}$ into two
copies of the reals. A direct counterexample to
$(\ref{eq:ym_ant})\Rightarrow (\ref{eq:ym_cons})$ of
Proposition~\ref{prop:ym-eq-interpol} can be obtained by taking
$\delta_i(x, y_i)\equiv y_i+y_i= x$ for $i=1,2$ so that $v(x)\equiv
{1\over 2}*x$ in (\ref{eq:ym_cons}) and the function symbol ${1\over
  2}*\_$ is required.

For \emph{non-convex} theories, the notion of equality interpolation
in this paper is strictly more general than the one proposed in the
extended version of~\cite{ym}.  Such a notion, to be called \emph{YM
  equality interpolating} below, requires quantifier-free
interpolation and the following condition:

\noindent $-$ for every tuples $\ux$, $\uz_1$, $\uz_2$ of variables,
further tuples $\uy_1=y_{11}, \dots, y_{1n}$, $\uy_2=y_{21}, \dots,
y_{2n}$ of variables, and pairs $\delta_1(\ux,\uz_1, \uy_1),
\delta_2(\ux, \uz_2, \uy_2)$ of 
conjunctions of literals,
\begin{equation*}
  \text{if }
  \delta_1(\ux, \uz_1,\uy_1)\wedge
  \delta_2(\ux, \uz_2, \uy_2)\vdash_T
  \bigvee_{i=1}^n (y_{1i}=y_{2i}) \text{ holds,}
\end{equation*}
then there exists a tuple $\uv(\ux)=v_1, \dots, v_n$ of terms such
that
\begin{equation*}
  \delta_1(\ux,\uz_1, \uy_1)\wedge
  \delta_2(\ux,\uz_2, \uy_2)\vdash_T
  \bigvee_{i=1}^n (y_{1i} = v_i \wedge v_i = y_{2i}).
\end{equation*} 
We show that the notion of YM equality interpolation implies that of
equality interpolation proposed in this paper.  Indeed, if a convex
theory is YMc equality interpolating, then it is also YM equality
interpolating.  Since $\EUF(\Sigma)$ is convex and YMc equality
interpolating (as shown in~\cite{ym}), it is YM equality
interpolating.  By Theorems~\ref{prop:strong_amalgamation_needed}
and~\ref{thm:strong_amalgamation_syntactic}
(and the combination result of~\cite{ym}), if a theory $T$ is YM
equality interpolating, it is also equality interpolating in the sense
of Definition~\ref{def:eq-interpol}.  The converse does not hold,
i.e.\ our notion is \emph{strictly weaker} than YM equality
interpolation.  To prove this, we define a (non-convex) theory
$T_{cex}$ that has the strong sub-amalgamation property but is not YM
equality interpolating.  Let the signature of $T_{cex}$ contain three
propositional letters $p_1, p_2$ and $p_3$, three constant symbols
$c_1, c_2$, and $c_3$, and a unary predicate $Q$.  $T_{cex}$ is
axiomatized by the following sentences: exactly one among $p_1, p_2$
and $p_3$ holds, $c_1, c_2$, and $c_3$ are distinct, $Q(x)$ holds for
no more than one $x$, and $p_i\to Q(c_i)$ for $i=1, 2, 3$.  It is easy
to see that $T_{cex}$ is stably infinite and has the strong
sub-amalgamation property
($T_{cex}$ is non-convex since $Q(x)\wedge y_1=c_1\wedge y_2=c_2\wedge
y_3=c_3$ implies the disjunction $x=y_1\vee x=y_2\vee x=y_3$ without
implying any single disjunct).  Now, notice that $Q(x)\wedge
Q(y)\vdash_{T_{cex}} x=y$.  According to the definition of the YM
equality interpolating property (see above), there should be a
\emph{single} ground term $v$ such that $Q(x)\wedge
Q(y)\vdash_{T_{cex}} x=v \wedge y=v$.  This cannot be the case since
we must choose among one of the three constants $c_1, c_2$, $c_3$ to
find such a term $v$ and none of these choices fits our
purposes. Hence, $T_{cex}$ is not YM equality interpolating although
it has the strong sub-amalgamation property and hence it is equality
interpolating according to Definition~\ref{def:eq-interpol}.

To conclude the comparison with~\cite{ym}, since the notion of
equality interpolation of this paper is \emph{strictly weaker} than
that of YM equality interpolation, the scope of applicability of our
result about the modularity of theories admitting quantifier-free
interpolation (i.e.\ Corollary~\ref{coro:sub-amalgmation-sufficient}
above) is 
\emph{broader} 
than the one in the extended version
of~\cite{ym}.

\section{An interpolation algorithm for combinations of theories}
\label{sec:algorithms}

Although the notion of equality interpolation toghether with
Corollary~\ref{coro:sub-amalgmation-sufficient} allow us to establish
the quantifier-free interpolation for all those theories obtained by
combining a theory axiomatizing a container data structure (such as
\EUF, $\mathcal{RDS}$, $\mathcal{LST}$, or $\AXDIFF$) with relevant
fragments of Arithmetics (such as $\mathcal{LAR}$, $\mathcal{IDL}$,
$\mathcal{UTVPI}$, or $\mathcal{LAI}$), just knowing that
quantifier-free interpolants exist may not be sufficient.  It would be
desirable to compute interpolants for combinations of theories by
modularly reusing the available interpolation algorithms for the
component theories.  This is the subject of this section.

To simplify the technical development, we work with ground formulae
over signatures expanded with free constants instead of quantifier
free formulae as done in the previous sections.  We use the letters
$A, B, \dots$ to denote finite sets of ground formulae; the logical
reading of a set of formulae is the conjunction of its elements.  For
a signature $\Sigma$ and set $A$ of formulae, $\Sigma^A$ denotes the
signature $\Sigma$ expanded with the free constants occurring in $A$.
Let $A$ and $B$ be two finite sets of ground formulae in the
signatures $\Sigma^A$ and $\Sigma^B$, respectively, and $\Sigma^C :=
\Sigma^A \cap \Sigma^B$.  Given a term, a literal, or a formula
$\varphi$ we call it:
\begin{itemize}
\item {\em \abcommon} iff it is defined over $\Sigma^C$;
\item {\em \alocal} (resp. {\em \blocal}) if it is defined over
  $\Sigma^A$ (resp. $\Sigma^B$);
\item {\em $A$-strict} (resp. {\em $B$-strict}) iff it is \alocal
  (resp. \blocal) but not \abcommon;
\item {\em \abmixed} if it contains symbols in both $(\Sigma^A
  \setminus \Sigma^C)$ and $(\Sigma^B \setminus \Sigma^C)$;
\item {\em \abpure} if it does not contain symbols in both $(\Sigma^A
  \setminus \Sigma^C)$ and $(\Sigma^B \setminus \Sigma^C)$.
\end{itemize}
(Sometimes in the literature about interpolation, ``\alocal'' and
``\blocal'' are used to denote what we call here ``$A$-strict'' and
``$B$-strict''). 

\subsection{Interpolating metarules}
\label{subsec:metarules}
Our combined interpolation method is based on the abstract framework
introduced in~\cite{RTA} (to which, the interested reader is pointed
for more details) and used also in~\cite{zfrocos} that is based on 
 `metarules.'  
A metarule applies (bottom-up) to a pair $A,B$ of finite sets of ground
formulae\footnote{In~\cite{RTA,zfrocos}, metarules manipulate 
  pairs of finite sets of literals instead of ground formulae; 
  the difference is immaterial.} producing an equisatisfiable pair of
sets of formulae.  Each metarule comes with a proviso for its
applicability and an instruction for the computation of the
interpolant.  As an example, consider the metarule (Define0):
\begin{center}
\begin{tabular}{ccl}
  $A\cup\{a=t\} \sep B\cup\{a=t\}$ 
  & \quad\quad &
  \emph{Proviso}: $t$ is $AB$-common, $a$ is fresh \\ \cline{1-1}
  $A \sep B$ 
  & \quad\quad &
  \emph{Instruction}: $\phi' \coincide \phi(t/a)$.
\end{tabular}
\end{center}
It is 
not difficult to see 
that the $A\cup B$ is equisatisfiable to
$A\cup B\cup\{ a=t\}$ since $a$ is a fresh variable that has been
introduced to re-name the $AB$-common term $t$ according to the
proviso of (Define0).  The instruction attached to (Define0) allows
for the computation of the interpolant $\phi'$ by eliminating the fresh
constant $a$ from the recursively known interpolant $\phi$.
The idea is to build an \emph{interpolating metarules refutation} 
for a given unsatisfiable
$A_0\cup B_0$, i.e.\ a labeled tree having the following properties: (i) nodes
are labeled by pairs of finite sets of ground formulae; (ii) the root
is labeled by $A_0, B_0$; (iii) the leaves are labeled by a pair $\tilde
A, \tilde B$ such that $\bot \in \tilde A\cup \tilde B$; (iv) each
non-leaf node is the conclusion of a metarule
and its successors are the premises of that metarule (the complete
list of metarules is in Appendix~\ref{app:metarules}).  Once an
interpolating metarules refutation has been built, it is possible to
recursively compute the interpolant by using (top-down) the
instructions attached to the metarules in the tree:

\begin{proposition}[\hspace{.5pt}\cite{RTA}]
  \label{prop:metarules}
  If there exists an interpolating metarules refutation for $A_0, B_0$ then there
  is a quan\-ti\-fier-free interpolant for $A_0, B_0$ (i.e., there exists
  a quan\-ti\-fier-free \abcommon sentence $\phi$ such that $A_0\vdash
  \phi$ and $B_0 \wedge \phi \vdash \bot$). The interpolant $\phi$ is
  recursively computed by applying the relevant interpolating
  instructions of the metarules. 
\end{proposition}
The idea to design the combination algorithm is the following.  We
design transformations instructions that can be non-deterministically
applied to a pair $A_0, B_0$. Each of the transformation instructions
is \emph{justified by metarules}, in the sense that it is just a
special sequence of applications of metarules.
The instructions are such that, whenever they are applied exhaustively
to a pair such that $A_0\cup B_0$ is unsatisfiable, they produce a
tree which is an interpolating metarules refutation for $A_0, B_0$
from which an interpolant can be extracted according to 
Proposition~\ref{prop:metarules}.

\subsection{A quantifier-free interpolating algorithm}
\label{subsec:combined-interpolating-algo}

Let $T_i$ be a stably-infinite and equality interpolating theory over
the signature $\Sigma_i$ such that
the $SMT(T_i)$ problem is decidable
and $\Sigma_1\cap \Sigma_2=\emptyset$ (for $i=1,2$).  We assume the
availability of algorithms for $T_1$ and $T_2$ that are able not only
to compute quantifier-free interpolants but also the tuples $\uv$ of
terms in
Definition~\ref{def:eq-interpol} for equality interpolation.  Since
the $SMT(T_i)$ problem is decidable for $i=1,2$, it is always possible
to build an equality interpolating algorithm by enumeration;
in practice, better 
algorithms 
can
be designed 
(see ~\cite{ym} for \EUF, $\mathcal{LST}$, $\mathcal{LAR}$
and Appendix~\ref{app:proofs} for the possibility to use  quantifier elimination 
to this aim).
 
Let $\Sigma:= \Sigma_1\cup \Sigma_2$, $T:=T_1\cup T_2$, and $A_0, B_0$
be a $T$-unsatisfiable pair of finite sets of ground formulae over the
signature $\Sigma^{A_0\cup B_0}$.  Like in the Nelson-Oppen
combination method, 
we have a pre-processing step in which
we {purify} $A_0$ and $B_0$ so as 
to eliminate from them the literals which are neither $\Sigma_1$- nor $\Sigma_2$-literals.
To do this, it is sufficient to repeatedly apply
the technique of ``renaming terms by constants'' described below.
Take a term $t$ (occurring in a literal from $A_0$ or from  $B_0$), add the equality $a=t$ for a fresh constant $a$  and replace 
all the occurrences of $t$ by $a$. 
The transformation can be
justified by the following sequence of metarules: Define1, Define2,
Redplus1, Redplus2, Redminus1, Redminus2.  For example, in the case of
the renaming of some term $t$ in $A_0$, the metarule Define1 is used
to add the explicit definition $a=t$ to $A_0$, the metarule Redplus1
to add the formula $\phi(a/t)$ for each $\phi\in A_0$, and the
metarule Redminus1 to remove from $A_0$ all the \formulae $\phi$ in which $t$ occurs (except $a=t$).

Because of purification, from now on, \emph{we assume to manipulate
  pairs $A, B$ of sets of ground \formulae where literals built up of only  $\Sigma_1$- or of only $\Sigma_2$-symbols
occur} 
(besides free constants): this invariant will be in fact maintained during the execution of our algorithm.
Given such a pair $A, B$, we denote by
$A_1$ and $A_2$  the subsets of $\Sigma_1^A$- and
$\Sigma_2^A$-formulae belonging to $A$; the sub-sets $B_1$ and $B_2$ of
$B$ are defined similarly.  Notice that \emph{it is false} that
$A\coincide A_1\cup A_2$ and $B\coincide B_1\cup B_2$, since
quantifier-free formulae can mix $\Sigma_1$- and $\Sigma_2$-symbols
even if the literals they are built from do not.
%

Before presenting our interpolation algorithm for the combination of
theories, we need to import a technique, called \emph{Term Sharing},
from~\cite{RTA}.  Suppose that $A$ contains a literal $a =t$, where
the term $t$ is \abcommon and the free constant $a$ is \astrict (a
symmetric technique applies to $B$ istead of $A$). Then it is possible
to ``make $a$ \abcommon'' in the following way. First, introduce a
fresh \abcommon constant $c$ with the explicit definition $c=t$ (to be
inserted both in $A$ and in $B$, as justified by metarule (Define0));
then replace the literal $a=t$ by $a=c$ and replace $a$ by $c$
everywhere else in $A$; finally, delete $a=c$ too.  The result is a
pair $(A, B)$ where basically nothing has changed but $a$ has been
renamed to an \abcommon constant $c$ (the
transformation can be 
easily
justified by a suitable subset of the metarules).

An \emph{$A$-relevant atom} is either an atomic formula occurring in
$A$ or it is an \alocal equality between free constants; an
\emph{$A$-assignment} is a Boolean assignment $\alpha$ to relevant
$A$-atoms satisfying $A$, seen as a set of propositional \formulae
(relevant $B$-atoms and $B$-assignements are defined similarly).
Below, we use the notation $\alpha$ to denote both the assignement
$\alpha$ and the set of literals $\{L\,\vert\,
\alpha(L)=\mathit{true}\}$.

 We are now in the position
to present the collection of
transformations
that should be applied non-deterministically and
exhaustively to a pair of purified sets of ground \formulae
(all the transformations below can be justified by
metarules, the justification is  straightforward and
left to the reader). 
In the following, let $i\in \{1,2\}$ and $X\in \{A, B\}$.
\begin{description}
%
\item[\textbf{Terminate}$_i$:] if $A_i\cup B_i$ is
  $T_i$-unsatisfiable and $\bot\not \in A\cup B$, use the interpolation algorithm for $T_i$ to
  find a ground \abcommon $\theta$ such that $A_i\vdash_{T_i}\theta$ and
  $\theta\wedge B_i\vdash_{T_i}\bot$; then add $\theta$ and $\bot$ to $B$.
\item[\textbf{Decide}$_X$:] 
if there is no $X$-assignment $\alpha$ such that $\alpha\subseteq X$, pick one of them (if there are none, add $\bot$ to $X$); 
then update $X$ to $X\cup \alpha$.
%
\item[\textbf{Share}$_i$:] let $\ua=a_1, \dots, a_n$ be the tuple of
  the current \astrict free constants and 
  $\ub=b_1, \dots, b_m$ be the tuple of the current \bstrict free
  constants.  Suppose that $A_i\cup B_i$ is $T_i$-satisfiable, but
  $A_i\cup B_i \cup \{\ua\cap \ub=\emptyset\}$ is 
  $T_i$-unsatisfiable.
  Since $T_i$ is equality interpolating, there must exist \abcommon
  $\Sigma_i$-ground terms $\uv\coincide v_1, \dots, v_p$ such that
  \begin{eqnarray*}
    A_i\cup B_i \vdash_{T_i} 
    (\ua\cap \uv \not= \emptyset) \vee (\ub\cap \uv \not= \emptyset).
  \end{eqnarray*}
  Thus the union of $A_i\cup \{\ua\cap \uv = \emptyset\}$ and of 
   $B_i\cup \{\ub\cap \uv = \emptyset\}$ is not $T_i$-satisfiable and
  invoking the available interpolation algorithm for $T_i$, we can
  compute a ground \abcommon $\Sigma_i$-formula $\theta$ such that $A\vdash_{T_i} \theta
  \vee \ua\cap \uv\not=\emptyset$ and $\theta\wedge B \vdash_{T_i}
  \ub\cap \uv\not=\emptyset$.  We choose among $n*p+m*p$
  alternatives in order to non-deterministically update $A, B$.  For
  the first $n*p$ alternatives, we add some $a_i=v_j$ (for $1\leq
  i\leq n$, $1\leq j\leq p$) to $A$.  For the last $m*p$ alternatives,
  we add $\theta$ to $A$ and some $\{\theta, b_i=v_j\}$ to $B$ (for
  $1\leq i\leq m$, $1\leq j\leq p$).  Term sharing
   is finally applied to the updated pair in order
  to decrease the number of the \astrict or \bstrict free
  constants.
\end{description}
Let $\mathsf{CI}(T_1,T_2)$ be the procedure that, once run on an
unsatifiable pair $A_0, B_0$, first purifies it, then
non-deterministically and exhaustively applies the transformation
rules above, and finally extracts an interpolant by using the
instructions associated to the metarules.
\begin{theorem}
  \label{thm:main}
  Let $T_1$ and $T_2$ be two signature disjoint, stably-infinite, and
  equality interpolating theories having decidable SMT problems.
  Then, $\mathsf{CI}(T_1,T_2)$ is a quantifier-free interpolation
  algorithm for the combined theory $T_1\cup T_2$.
\end{theorem}
Algorithm $\mathsf{CI}(T_1,T_2)$ paves the way to reuse
quantifier-free interpolation algorithms for both conjunctions (see,
e.g.,~\cite{rybalchenko-stokkermans}) or arbitrary Boolean
combinations of literals (see, e.g.,~\cite{CGS08}).  In particular,
the capability of reusing interpolation algorithms that can
efficiently handle the Boolean structure of formulae seems to be key
to enlarge the scope of applicability of verification methods based on
interpolants~\cite{mcmillan-z3-interpolation}.  Indeed, one major
issue to address to make $\mathsf{CI}(T_1,T_2)$ practically usable is
to eliminate the non-determinism.  We believe this is possible by
adapting the Delayed Theory Combination approach~\cite{BBC+05c}.  


\section{Conclusion and Related Work}
\label{sec:conclusion}
The results of this paper cover several results for the
quantifier-free interpolation of combinations of theories that are
known from the literature, e.g.,
\EUF and $\mathcal{LST}$~\cite{ym}, \EUF and
$\mathcal{LAR}$~\cite{McM05,CGS08,rybalchenko-stokkermans},
\EUF and $\mathcal{LAI}$~\cite{brillout}, $\mathcal{LST}$ with
$\mathcal{LAR}$~\cite{ym}, and $\AXDIFF$ with
$\mathcal{IDL}$~\cite{zfrocos}.  To the best of our knowledge, the
quantifier-free interpolation of the following combinations are new:
(a) $\mathcal{RDS}$ with $\mathcal{LAR}$, $\mathcal{IDL}$,
$\mathcal{UTVPI}$, $\mathcal{LAI}$, and $\AXDIFF$, (b) $\mathcal{LST}$
with $\mathcal{IDL}$, $\mathcal{UTVPI}$, $\mathcal{LAI}$, and
$\AXDIFF$, and (c) $\AXDIFF$ with $\mathcal{LAR}$, $\mathcal{UTVPI}$,
and $\mathcal{LAI}$.

In Section~\ref{subsec:comparison}, we have extensively discussed the
closely related work of~\cite{ym}, where the authors illustrate a
method to derive interpolants in a Nelson-Oppen combination procedure,
provided that the component theories satisfy certain hypotheses.  The
work in~\cite{bonacina2}, among other contributions, recasts the
method of~\cite{ym} in the context of the $DPLL(T)$ paradigm.  An
alternative combination method is in~\cite{goel-sava-tinelli} that has
been designed to be efficiently incorporated in state-of-the-art SMT
solvers but is complete only for convex theories.
An interpolating theorem prover is described in~\cite{McM05}, where a
sequent-like calculus is used to derive interpolants from proofs in
propositional logic, equality with uninterpreted functions, linear
rational arithmetic, and their combinations.  The ``split'' prover
in~\cite{jhala} applies a sequent calculus for the synthesis of
interpolants along the lines of that in~\cite{McM05} and is tuned for
predicate abstraction.  The ``split'' prover can handle combinations
of theories involving that of arrays without extensionality and
fragments of Linear Arithmetic.  The \textsc{CSIsat}~\cite{csisat}
permits the computation of quantifier-free interpolants over a
combination of \EUF and $\mathcal{LAR}$ refining the combination
method in~\cite{ym}.  A version of \textsc{MathSAT}~\cite{CGS08}
features interpolation capabilities for 
\EUF,
$\mathcal{LAR}$, $\mathcal{IDL}$, $\mathcal{UTVPI}$ and $\EUF+\mathcal{LAR}$ by extending
 Delayed Theory Combination~\cite{BBC+05c}.  Theorem~\ref{thm:main}
is the key to combine the strength of these tools and to widen the
scope of applicability of available interpolation algorithms to richer
combinations of theories.  
Methods~\cite{KMZ06,voronkov,brillout,mcmillan-z3-interpolation} for
the computation of quantified interpolants in the combination of the
theory of arrays 
and Presburger Arithmetic have been proposed.
Our work focus on quantifier-free interpolants by identifying suitable
variants of the component theories (e.g., \AXDIFF instead of \AXEXT and
$\mathcal{LAI}$ instead of Presburger Arithmetic).  Orthogonal to our
approach is the work in~\cite{stokkermans} where interpolation
algorithm are developed for extensions of convex theories admitting
quantifier-free interpolation.

The framework proposed in this paper allows us to give a uniform and
coherent view of many results available in the literature and we hope
that it will be  the starting point for new developements.

\bibliographystyle{plain}
\bibliography{brutt,ghila,ranis}

 \newpage
\appendix

\section{List of Metarules}
\label{app:metarules}

\begin{table}[h!]
\scalebox{.95}{
\centering

\begin{tabular}{|cccccccccccc|}

\hline

\multicolumn{3}{|c|}{Close1} &
\multicolumn{3}{|c|}{Close2} &
\multicolumn{3}{|c|}{Propagate1} &
\multicolumn{3}{|c|}{Propagate2} \\

\hline


\multicolumn{3}{|c|}{
\begin{minipage}{.25\textwidth}
\begin{center}
\vspace{8pt}
\begin{prooftree}
\AxiomC{$~$}
\UnaryInfC{$A \sep B$}
\end{prooftree}
\vspace{8pt}
{\scriptsize
\begin{tabular}{rl}
\emph{Prov.}: & $A$ is unsat. \\
\emph{Instr.}: & $\phi' \coincide \bot$.
\end{tabular}
}
\vspace{8pt}
\end{center}
\end{minipage}
}

&


\multicolumn{3}{|c|}{
\begin{minipage}{.25\textwidth}
\begin{center}
\vspace{8pt}
\begin{prooftree}
\AxiomC{$~$}
\UnaryInfC{$A \sep B$}
\end{prooftree}
\vspace{8pt}
{\scriptsize
\begin{tabular}{rl}
\emph{Prov.}: & $B$ is unsat. \\
\emph{Instr.}: & $\phi' \coincide \top$.
\end{tabular}
}
\vspace{8pt}
\end{center}
\end{minipage}
}

&


\multicolumn{3}{|c|}{
\begin{minipage}{.25\textwidth}
\begin{center}
\vspace{8pt}
\begin{prooftree}
\AxiomC{$A \sep B\cup\{\psi\}$}
\UnaryInfC{$A \sep B$}
\end{prooftree}
\vspace{8pt}
{\scriptsize
\begin{tabular}{r@{ }l}
\emph{Prov.}: & $A\vdash \psi$ and \\ 
              & $\psi$ is $AB$-common. \\
\emph{Instr.}: & $\phi' \coincide \phi \wedge \psi$.
\end{tabular}
}
\vspace{8pt}
\end{center}
\end{minipage}
}

&


\multicolumn{3}{|c|}{
\begin{minipage}{.25\textwidth}
\begin{center}

\begin{prooftree}
\AxiomC{$A\cup\{\psi\} \sep B$}
\UnaryInfC{$A \sep B$}
\end{prooftree}

\vspace{8pt}
{\scriptsize
\begin{tabular}{r@{ }l}
\emph{Prov.}: & $B\vdash \psi$ and \\
              & $\psi$ is $AB$-common. \\
\emph{Instr.}: & $\phi' \coincide \psi \to\phi$.
\end{tabular}
}
\end{center}
\end{minipage}
}

\\
\hline
\hline

\multicolumn{4}{|c|}{Define0} &
\multicolumn{4}{|c|}{Define1} &
\multicolumn{4}{|c|}{Define2} \\

\hline


\multicolumn{4}{|c|}{
\begin{minipage}{.33\textwidth}
\begin{center}
\vspace{8pt}
\begin{prooftree}
\AxiomC{$A\cup\{a=t\} \sep B\cup\{a=t\}$}
\UnaryInfC{$A \sep B$}
\end{prooftree}
\vspace{8pt}
{\scriptsize
\begin{tabular}{r@{ }l}
\emph{Prov.}: & $t$ is $AB$-common, $a$ fresh. \\
\emph{Instr.}: & $\phi' \coincide \phi(t/a)$.
\end{tabular}
}
\vspace{8pt}
\end{center}
\end{minipage}
}

&


\multicolumn{4}{|c|}{
\begin{minipage}{.33\textwidth}
\begin{center}
\begin{prooftree}
\AxiomC{$A\cup\{a=t\} \sep B$}
\UnaryInfC{$A \sep B$}
\end{prooftree}

\vspace{8pt}

{\scriptsize
\begin{tabular}{r@{ }l}
\emph{Prov.}: & $t$ is \alocal and $a$ is fresh. \\
\emph{Instr.}: & $\phi' \coincide \phi$.
\end{tabular}
}
\end{center}
\end{minipage}
}

&


\multicolumn{4}{|c|}{
\begin{minipage}{.33\textwidth}
\begin{center}
\begin{prooftree}
\AxiomC{$A \sep B\cup\{a=t\}$}
\UnaryInfC{$A \sep B$}
\end{prooftree}

\vspace{8pt}

{\scriptsize
\begin{tabular}{rl}
\emph{Prov.}: & $t$ is \blocal and $a$ is fresh. \\
\emph{Instr.}: & $\phi' \coincide \phi$.
\end{tabular}
}
\end{center}
\end{minipage}
}

\\
\hline
\hline

\multicolumn{6}{|c|}{Disjunction1} &
\multicolumn{6}{|c|}{Disjunction2} \\

\hline


\multicolumn{6}{|c|}{
\begin{minipage}{.5\textwidth}
\begin{center}
\vspace{8pt}
\begin{prooftree}
 \AxiomC{$\cdots ~~~A\cup\{\psi_k\} \sep B~~~\cdots~~~$}                                              
%
\UnaryInfC{$A \sep B$}                                                                                
\end{prooftree}
\vspace{8pt}
{\scriptsize
\begin{tabular}{rl}
\emph{Prov.}: & $\bigvee_{k=1}^n\psi_k$ is \alocal and $A \vdash \bigvee_{k=1}^n\psi_k$. \\
%
\emph{Instr.}: & $\phi' \coincide \bigvee_{k=1}^n\phi_k$.
\end{tabular}
}
\vspace{8pt}
\end{center}
\end{minipage}
}

&


\multicolumn{6}{|c|}{
\begin{minipage}{.5\textwidth}
\begin{center}
\vspace{8pt}
\begin{prooftree}
\AxiomC{$\cdots ~~~A \sep B\cup\{\psi_k\}~~~\cdots~~~$}                     
%
\UnaryInfC{$A \sep B$}                                                       
\end{prooftree}
\vspace{8pt}
{\scriptsize
\begin{tabular}{rl}
\emph{Prov.}: & $\bigvee_{k=1}^n\psi_k$ is \blocal and $B \vdash \bigvee_{k=1}^n\psi_k$. \\
\emph{Instr.}: & $\phi' \coincide \bigwedge_{k=1}^n\phi_k$.
\end{tabular}
}
\vspace{8pt}
\end{center}
\end{minipage}
}

\\
\hline
\hline

\multicolumn{3}{|c|}{Redplus1} &
\multicolumn{3}{|c|}{Redplus2} &
\multicolumn{3}{|c|}{Redminus1} &
\multicolumn{3}{|c|}{Redminus2} \\

\hline


\multicolumn{3}{|c|}{
\begin{minipage}{.25\textwidth}
\begin{center}
\vspace{8pt}
\begin{prooftree}
\AxiomC{$A \cup\{\psi\}\sep B$}
\UnaryInfC{$A \sep B$}
\end{prooftree}
\vspace{8pt}
{\scriptsize
\begin{tabular}{rl}
\emph{Prov.}: & $A\vdash \psi$ and \\ 
              & $\psi$ is \alocal. \\
\emph{Instr.}: & $\phi' \coincide \phi$.
\end{tabular}
}
\vspace{8pt}
\end{center}
\end{minipage}
}

&


\multicolumn{3}{|c|}{
\begin{minipage}{.25\textwidth}
\begin{center}
\vspace{8pt}
\begin{prooftree}
\AxiomC{$A \sep B\cup\{\psi\}$}
\UnaryInfC{$A \sep B$}
\end{prooftree}
\vspace{8pt}
{\scriptsize
\begin{tabular}{rl}
\emph{Prov.}: & $B\vdash \psi$ and \\ 
              & $\psi$ is \blocal. \\
\emph{Instr.}: & $\phi' \coincide \phi$.
\end{tabular}
}
\vspace{8pt}
\end{center}
\end{minipage}
}

&


\multicolumn{3}{|c|}{
\begin{minipage}{.25\textwidth}
\begin{center}
\vspace{8pt}
\begin{prooftree}
\AxiomC{$A \sep B$}
\UnaryInfC{$A \cup\{\psi\}\sep B$}
\end{prooftree}
\vspace{8pt}
{\scriptsize
\begin{tabular}{rl}
\emph{Prov.}: & $A\vdash \psi$ and \\
              & $\psi$ is \alocal. \\
\emph{Instr.}: & $\phi' \coincide \phi$.
\end{tabular}
}
\vspace{8pt}
\end{center}
\end{minipage}
}

&


\multicolumn{3}{|c|}{
\begin{minipage}{.25\textwidth}
\begin{center}
\vspace{8pt}
\begin{prooftree}
\AxiomC{$A \sep B$}
\UnaryInfC{$A \sep B\cup\{\psi\}$}
\end{prooftree}
\vspace{8pt}
{\scriptsize
\begin{tabular}{rl}
\emph{Prov.}: & $B\vdash \psi$ and \\ 
              & $\psi$ is \blocal. \\
\emph{Instr.}: & $\phi' \coincide \phi$.
\end{tabular}
}
\vspace{8pt}
\end{center}
\end{minipage}
}
\\

\hline
\hline

\multicolumn{4}{|c|}{ConstElim1} &
\multicolumn{4}{|c|}{ConstElim2} &
\multicolumn{4}{|c|}{ConstElim0} \\

\hline


\multicolumn{4}{|c|}{
\begin{minipage}{.33\textwidth}
\begin{center}
\vspace{8pt}
\begin{prooftree}
\AxiomC{$A \sep B$}
\UnaryInfC{$A\cup\{a=t\}\sep B$}
\end{prooftree}
\vspace{8pt}
{\scriptsize
\begin{tabular}{rl}
\emph{Prov.}: & $a$ is $A$-strict and \\ 
              & does not occur in $A, t$. \\
\emph{Instr.}: & $\phi' \coincide \phi$.
\end{tabular}
}
\vspace{8pt}
\end{center}
\end{minipage}
}

&


\multicolumn{4}{|c|}{
\begin{minipage}{.33\textwidth}
\begin{center}
\vspace{8pt}
\begin{prooftree}
\AxiomC{$A \sep B$}
\UnaryInfC{$A \sep B\cup\{b=t\}$}
\end{prooftree}
\vspace{8pt}
{\scriptsize
\begin{tabular}{rl}
\emph{Prov.}: & $b$ is $B$-strict and \\ 
              & does not occur in $B, t$. \\
\emph{Instr.}: & $\phi' \coincide \phi$.
\end{tabular}
}
\vspace{8pt}
\end{center}
\end{minipage}
}

&


\multicolumn{4}{|c|}{
\begin{minipage}{.33\textwidth}
\begin{center}
\vspace{8pt}
\begin{prooftree}
\AxiomC{$A \sep B$}
\UnaryInfC{$A\cup\{c=t\} \sep B\cup\{c=t\}$}
\end{prooftree}
\vspace{8pt}
{\scriptsize
\begin{tabular}{r@{ }l}
\emph{Prov.}: & $c$, $t$  are \abcommon, \\ 
              & $c$ does not occur in $A, B, t$. \\
\emph{Instr.}: & $\phi' \coincide \phi$.
\end{tabular}
}
\vspace{8pt}
\end{center}
\end{minipage}
}

\\

\hline

\end{tabular}
}

\bigskip
\caption{\footnotesize{Interpolating Metarules: 
each rule has a proviso $Prov.$ and an instruction $Instr.$ for
recursively computing the new interpolant $\phi'$ 
 from the old
one(s) $\phi, \phi_1, \dots, \phi_k$.
Metarules are applied \emph{bottom-up} and interpolants are computed
\emph{top-down}.
  }
}
\label{tab:metarules}
\end{table}


\newpage

\section{Proofs}
\label{app:proofs}

We give here all the proofs not included in the text.

\subsection{Proofs for Section~\ref{sec:strong_amalagamation}}

\begin{lemma}
  \label{lem:app}
  Let $T$ be a theory in a signature $\Sigma$ and let $\ua, \ub, \uc$
  be tuples of (distinct) free constants; let also $\Theta_1,
  \Theta_2$ be sets of ground formulae having the following
  properties:
 \begin{description}
  \item[-] in $\Theta_1$ at most the free constants $\ua, \uc$ occur;
  \item[-] in $\Theta_2$ at most the free constants $\ub, \uc$ occur;
  \item[-] there is no ground formula $\theta(\uc)$ such that $\Theta_1\vdash_T \theta(\uc)$ and $\Theta_2\vdash_T \neg\theta(\uc)$.
 \end{description}
Then there are models $\cM_1, \cM_2$ of $T$ such that $\cM_1\models \Theta_1$, 
$\cM_2\models\Theta_2$ and such that the intersection of the supports of $\cM_1$ and $\cM_2$ is precisely the substructure generated by the interpretation of the constants $\uc$.
\end{lemma}

\begin{proof}
Let us call $\Sigma^A$ the signature $\Sigma$ expanded with the
free constants $\ua\cup\uc$ and  $\Sigma^B$ the signature $\Sigma$ expanded
with the free constants  $\ub\cup\uc$ (we put $\Sigma^C:= \Sigma^A\cap
\Sigma^B=\Sigma\cup \{\uc\}$).
As a first step, we build a maximal $T$-consistent set $\Gamma$ of
ground $\Sigma^A$-formulae and a maximal $T$-consistent set
$\Delta$ of ground $\Sigma^B$-formulae such that $\Theta_1\subseteq
\Gamma$, $\Theta_2\subseteq \Delta$, and $\Gamma\cap \Sigma^C = \Delta \cap
\Sigma^C$.\footnote{ By abuse, we use $\Sigma^C$ to indicate not only
  the signature $\Sigma^C$ but also the set of formulae in the
  signature $\Sigma^C$.  } For simplicity\footnote{ This is just to
  avoid a (straightforward indeed) transfinite induction argument.}
let us assume that $\Sigma$ is at most countable, so that we can fix
two enumerations
$$
\phi_1, \phi_2, \dots \qquad \psi_1, \psi_2, \dots
$$ 
of ground $\Sigma^A$- and $\Sigma^B$-formulae,
respectively. We build inductively $\Gamma_n, \Delta_n$ such that for
every $n$ (i) $\Gamma_n$ contains either $\phi_n$ or $\neg \phi_n$; (ii)
$\Delta_n$ contains either $\psi_n$ or $\neg \psi_n$; (iii) there is no
ground $ \Sigma^C$-formula $\theta$ such that $\Gamma_n \cup
\{\neg \theta\}$ and $\Delta_n\cup \{\theta\}$ are not $T$-consistent.  Once
this is done, we can get our $\Gamma, \Delta$ as $\Gamma\coincide\bigcup
\Gamma_n$ and $\Delta \coincide\bigcup \Delta_n$.

We let $\Gamma_0$ be $\Theta_1$ and $\Delta_0$ be $\Theta_2$ (notice that
(iii) holds by assumption). To build $\Gamma_{n+1}$ we have two
possibilities, namely $\Gamma_n \cup \{ \phi_n\}$ and $\Gamma_n \cup \{
\neg \phi_n\}$. Suppose they are both unsuitable because there are $\theta_1,
\theta_2\in \Sigma^C$ such that the sets
$$
\Gamma_n \cup \{ \phi_n,\neg \theta_1\}, \quad \Delta_n\cup \{\theta_1\}, \quad \Gamma_n \cup \{\neg \phi_n, \neg \theta_2\}, \quad \Delta_n\cup \{\theta_2\}
$$
are all $T$-inconsistent. If we put $\theta\coincide \theta_1\vee \theta_2$, we get that $\Gamma_n \cup \{\neg \theta\}$ and $\Delta_n\cup 
\{\theta\}$ are not $T$-consistent, contrary to induction hypothesis. A similar argument
shows that we can also build $\Delta_n$.

Let now $\cM_1$ be a model of $\Gamma$ and $\cM_2$ be a model of
$\Delta$. Consider the substructures $\cA_1, \cA_2$ of $\cM_1, \cM_2$
generated by the interpretations of the constants from $\Sigma^C$:
since they satisfy the same literals from $\Sigma^C$
(because  $\Gamma\cap \Sigma^C = \Delta \cap \Sigma^C$), we have that $\cA_1$ and $\cA_2$ are $\Sigma^C$-isomorphic.
 Up to renaming, we can suppose that $\cA_1$ and $\cA_2$ are just the same substructure. \myqed
\end{proof}

\vskip 2mm\noindent
\textbf{Theorem}~\ref{thm:interpolation-amalgamation} 
  \emph{A theory $T$ admits quantifier-free interpolants
  iff $T$ has the sub-amalgamation property.}

\begin{proof}  
\emph{Suppose first that $T$ has sub-amalgamation}; let $\phi, \psi$ be
quantifier-free formulae such that $\phi\wedge \psi$ is not
$T$-satisfiable. Let us replace variables with free constants in $\phi,
\psi$; let us call $\Sigma^A$ the signature $\Sigma$ expanded with the
free constants from $\phi$ and $\Sigma^B$ the signature $\Sigma$ expanded
with the free constants from $\psi$ (we put $\Sigma^C:= \Sigma^A\cap
\Sigma^B$). For reductio, suppose that there is no ground formula $\theta$
such that: (a) $\phi$ $T$-entails $\theta$; (b) $\theta\wedge \psi$ is
$T$-unsatisfiable; (c) only free constants from $\Sigma^C$ occur in
$\theta$. By Lemma~\ref{lem:app}, taking $\Theta_1:=\{\phi\}, \Theta_2:=\{\psi\}$, we know that there are 
models $\cM_1, \cM_2$ of $T$ such that $\cM_1\models \phi, \cM_2\models\psi$ and such that 
the intersection of the supports of $\cM_1$ and $\cM_2$ is precisely the substructure generated by the interpretation of the constants 
from $\Sigma^C$ (let us we call this substructure $\cA$ for short).
By the sub-amalgamation property, there is a $T$-amalgam $\cM$ of $\cM_1$ and $\cM_2$ over $\cA$. 
Now $\phi, \psi$ are ground formulae true in $\cM_1$ and $\cM_2$, respectively,
hence they are both true in $\cM$, which is impossible because $\phi\wedge \psi$ was assumed to be $T$-inconsistent.

\vskip 2mm
\emph{Suppose now that $T$ has quantifier-free interpolants}. Take two models $\cM_1$ and $ \cM_2$ of $T$ sharing a substructure $\cA$;
we can freely suppose (up to a renaming) that $\vert \cM_1\vert \cap \vert \cM_2\vert =\vert \cA\vert$ 
(we use the notation $\vert -\vert$ to indicated the support of a structure). 
 In order
to show that a $T$-amalgam of $\cM_1, \cM_2$ over $\cA$ exists, it is sufficient
(by Robinson Diagram Lemma~\cite{CK}) to show that $\Delta_{\Sigma}(\cM_1)\cup \Delta_{\Sigma}(\cM_2)$ is $T$-consistent,
where (for $i=1,2$) $\Delta_{\Sigma}(\cM_i)$  is the \emph{diagram} of $\cM_i$, namely the set of $\Sigma\cup \vert \cM_i\vert$-literals true in $\cM_i$.  

 If $\Delta_{\Sigma}(\cM_1)\cup \Delta_{\Sigma}(\cM_2)$ is not $T$-consistent, by the compactness theorem of first order logic, there exist a
$\Sigma\cup {\vert \cM_1\vert}$-ground sentence $\phi$ and a $\Sigma\cup {\vert \cM_2\vert}$-ground sentence $\psi$ 
such that (i) $\phi\wedge \psi$ is $T$-inconsistent; (ii) $\phi$ is a conjunction of literals from  $\Delta_{\Sigma}(\cM_1)$; (iii)
$\psi$ is a conjunction of literals from  $\Delta_{\Sigma}(\cM_2)$.
By the existence of quantifier-free interpolants, taking free constants instead of variables, 
we get that there exists a ground $\Sigma\cup \vert \cA\vert$-sentence $\theta$ such that $\phi$ $T$-entails $\theta$ and $\psi\wedge \theta$ is $T$-inconsistent. 
The former fact yields that $\theta$ is true in $\cM_1$ and hence also
in $\cA$ and in $\cM_2$,
because $\theta$
is ground.
 However, the fact that $\theta$ is true in $\cM_2$ contradicts the fact that $\psi\wedge \theta$ is $T$-inconsistent.
\myqed
\end{proof}

The following Lemma is part of the well-known  Nelson-Oppen combination 
results~\cite{TinHar-FROCOS-96},\cite{NO79}: 

\begin{lemma}\label{lem:NO}
Suppose that $T_1, T_2$ are two stably infinite theories in disjoint signatures $\Sigma_1, \Sigma_2$ and let $C$ be a set of free constants not belonging to $\Sigma_1\cup \Sigma_2$; let $\Gamma$ be a 
partition of $C$, i.e. a set of ground equalities or inequalities containing the literal $c_1 = c_2$ or the literal $c_1\neq c_2$, for all pairs of different constants from $C$.
For $i=1,2$, let $\Theta_i$ be a $T_i$-consistent set of ground $\Sigma_i\cup C$-formulae containing $\Gamma$. 
Then $\Theta_1\cup \Theta_2$ is $T_1\cup T_2$-consistent.  
\end{lemma}

\begin{proof}
Let $\cM_1, \cM_2$ be two models of $T_1\cup \Theta_1, T_2\cup \Theta_2$, respectively. By stable infiniteness and upward L\"ovenheim-Skolem theorem~\cite{CK}, we can assume that  
they are both infinite and have the same cardinality  (bigger than the cardinality of $C$). Thus there is a bijection $f$ among their supports and (as equalities of constants from $C$ are interpreted 
in the same way
 in $\cM_1$ and $\cM_2$) we can assume that $f(c^{\cM_1})=c^{\cM_2}$. Using this bijection, it is easy to lift the interpretation of the $\Sigma_2$-symbols from the support of $\cM_2$ to the support of
$\cM_1$. The lifted model is $\Sigma_2\cup C$-isomorphic to $\cM_2$, thus it is a model of $T_1\cup T_2\cup \Theta_1\cup \Theta_2$. \myqed
\end{proof}

\vskip 2mm\noindent
\textbf{Theorem}~\ref{thm:sub-amalgamation_transfer}\emph{
 Let $T_1, T_2$ be two stably infinite theories in disjoint signatures $\Sigma_1, \Sigma_2$.
 If $T_1, T_2$ both have the strong sub-amalgamation property, then so does 
$T_1\cup T_2$. 
}

\begin{proof}
 Consider two models $\cM_1, \cM_2$ of $T_1\cup T_2$ together with a common substructure $\cA$; 
we can freely suppose (up to a renaming) that $\vert \cM_1\vert \cap \vert \cM_2\vert =\vert \cA\vert$.  
By Robinson Diagram Lemma~\cite{CK}, it is sufficient to show the consistency
of $T_1\cup T_2\cup \Gamma_1\cup \Gamma_2$, where $\Gamma_i$  ($i=1,2$) is defined as
$$
\Gamma_i ~\coincide ~ \Delta_{\Sigma_i}(\cM_1)\cup \Delta_{\Sigma_i}(\cM_2)\cup \{ m_1\neq m_2~\vert~ m_1\in \vert \cM_1\vert\setminus \vert \cA\vert,
~ m_2\in \vert \cM_2\vert\setminus \vert \cA\vert\}   ~~.
$$ 
 By compactness, it is enough to show the $T_1\cup T_2$-consistency of the 
subset $T_1\cup T_2\cup \Gamma^0_1\cup \Gamma^0_2$
of $T_1\cup T_2\cup \Gamma_1\cup \Gamma_2$ mentioning just a finite set $C$ of free constants
from $\vert \cM_1\vert \cup \vert \cM_2\vert$. By the strong amalgamability of $T_1$ and $T_2$, we know that 
 $T_1\cup\Gamma^0_1$ and $T_2\cup\Gamma^0_2$ are both consistent. Now notice that for every pair $c_1, c_2$ of distinct constants  from $C$, the set $\Gamma_i$ 
(hence also the set $\Gamma_i^0$)
contains the negative literal $c_1\neq c_2$: in fact, this inequation is part of the definition of the diagram of a structure or (in case $c_1, c_2$ are from 
different supports) it has been added explicitly when building $\Gamma_1, \Gamma_2$. According to Lemma~\ref{lem:NO},
this is sufficient 
to infer the consistency of $T_1\cup T_2\cup \Gamma^0_1\cup \Gamma^0_2$, as $T_1, T_2$ are stably infinite.
\myqed
\end{proof}

\vskip 2mm\noindent
\textbf{Theorem}~\ref{prop:strong_amalgamation_needed} \emph{
Let $T$ be a theory admitting quantifier-free interpolation and let $\Sigma$ be a signature disjoint from the signature of $T$ and containing at least a unary predicate symbol.
  Then $T\cup \EUF(\Sigma)$ has quantifier-free interpolation iff $T$ has the strong sub-amalgamation property.
}

\begin{proof}
 (Below $\Sigma_T$ is the signature of $T$).
Let $T$ be strongly amalgamable and let $\cM_1, \cM_2$ be two models of $T\cup \EUF(\Sigma)$ sharing a submodel $\cM_0$
(as usual, we suppose that $\vert \cM_1\vert \cap \vert \cM_2\vert = \vert \cM_0\vert$). To amalgamate them, consider first a model $\cM$ of 
$T$ strongly amalgamating the $\Sigma_T$-reducts  of $\cM_1, \cM_2$ over the $\Sigma_T$-reduct of $\cM_0$. Since the amalgam is strong, 
up to isomorphism we can consider the support of
$\cM$ as a superset of $\vert \cM_1\vert \cup \vert \cM_2\vert$; thus it is easy to expand $\cM$ to a total structure interpreting  the symbols of $\Sigma$. The 
expansion is a model of $T\cup\EUF(\Sigma)$ amalgamating $\cM_1$ and $\cM_2$ over $\cM_0$.

Conversely, suppose that $T$ does not have the sub-amalgamation property.
 Let $\cM_1, \cM_2$ be models of $T_1$ and let $\cA$ be a substructure of them such that there are no data $\cM, \mu_1, \mu_2$ satisfying the conditions for the
strong sub-amalgamability property. This means that the set 
$$
\Gamma ~\coincide ~ \Delta_{\Sigma_1}(\cM_1)\cup \Delta_{\Sigma_1}(\cM_2)\cup \{ m_1\neq m_2~\vert~ m_1\in \vert \cM_1\vert\setminus \vert \cA\vert,
~ m_2\in \vert \cM_2\vert\setminus \vert \cA\vert\} 
$$ 
is not $T$-consistent. By compactness, there are 
$m^1_1, \dots m^k_1\in \vert \cM_1\vert\setminus \vert \cA\vert$ and 
$m^1_2, \dots m^k_2\in \vert \cM_2\vert\setminus \vert \cA\vert$
such that 
\begin{equation}
 \label{eq:counterexample}
T\cup \Delta_{\Sigma_1}(\cM_1)\cup \Delta_{\Sigma_1}(\cM_2) \models \bigvee_{j=1}^k m^j_1\uguale m^j_2  ~~.
\end{equation}
Expand now $\cM_1, \cM_2$ to $\Sigma_T\cup \Sigma$-structures as follows: 
the $\Sigma$-symbols are interpreted arbitrarily (but in such a way that $\cA$ remains a substructure of the expansions) apart from the unary predicate $P$,  
which is interpreted as the whole support of $\cM_1$
in the expansion of $\cM_1$ and as the support of $\cA$ in the expansion of $\cM_2$. From~\eqref{eq:counterexample}, it is then clear that sub-amalgamation
(hence quantifier-free interpolation) fails for $T_1\cup T_2$: in fact,  any $\cM\models T$ amalgamating $\cM_1, \cM_2$ over $\cA$, must identify 
some $m_1\in \vert \cM_1\vert\setminus \vert \cA\vert$
with some $m_2\in \vert \cM_2\vert\setminus \vert \cA\vert$, which is impossible as the interpretation of $P$ in $\cM$ must agree with the interpretations of 
$P$ in the expansions of $\cM_1$ and $\cM_2$. 
\myqed
\end{proof}

\subsection{Proofs for Section~\ref{sec:strong_amalagamation_syntactically}}
\label{subapp:proofs-of-sec:strong_amalagamation_syntactically}

Theorem~\ref{thm:strong_amalgamation_syntactic} shows the equivalence between strong amalgamability and equality interpolation; 
we add one equivalent characterization more in the statement below:

\vskip 2mm\noindent
\textbf{Theorem}~\ref{thm:strong_amalgamation_syntactic}\emph{
The following conditions are equivalent for a theory $T$ having quan\-ti\-fier-free interpolation:
\begin{description}
 \item[{\rm (i)}] $T$ is strongly sub-amalgamable;
 \item[{\rm (ii)}] $T$ is equality interpolating;
 \item[{\rm (iii)}] for every triple $\ux, \uy_1, \uy_2$ of tuples of variables and for every
pair of quantifier-free formulae $\delta_1(\ux, \uy_1), \delta_2(\ux, \uy_2)$
such that  
\begin{equation}
\label{eq:antecedent}
 \delta_1(\ux, \uy_1)\wedge \delta_2(\ux, \uy_2)\vdash_T \uy_1\cap \uy_2\neq \emptyset 
\end{equation}
there is a tuple $\uv(\ux)$ of terms such that
\begin{equation}
\label{eq:consequent}
 \delta_1(\ux, \uy_1)\wedge \delta_2(\ux, \uy_2)\vdash_T \uy_1\uy_2\cap \uv\neq \emptyset ~.
\end{equation}
\end{description}
}

\begin{proof} We first show ${\rm (i)}\Rightarrow {\rm (ii)}$. 
 Suppose first that $T$ is strongly sub-amalgamable; we show that~$\eqref{eq:antecedent_var}\Rightarrow\eqref{eq:consequent_var}$ holds by contraposition. So, 
let us fix tuples of fresh free constants $\ua, \um_1,\un_1, \um_2, \un_2$ and
suppose that 
for every finite tuple $\uv$ of $\Sigma\cup\{\ua\}$-ground terms, the  formula 
\begin{equation}\label{eq:consistent1}
\delta_1(\ua, \un_1,\um_1)\wedge \delta_2(\ua,\un_2 \um_2)\wedge (\um_1\um_2\cap \uv\uguale \emptyset) 
\end{equation}
is $T$-consistent (here $\Sigma$ is the signature of $T$). 
We claim that the set
\begin{equation}\label{eq:consistent2}
\{\delta_1(\ua, \un_1,\um_1), \delta_2(\ua,\un_2, \um_2)\}\cup \{\um_1\um_2\cap \uv\uguale \emptyset\}_{\uv} 
\end{equation}
is $T$-consistent, where
 $\uv$ varies over all possible tuples of such terms. In fact, if~\eqref{eq:consistent2} were not consistent, by compactness, there would be tuples 
of $\Sigma\cup\{\ua\}$-ground terms
$\uv_{1}, \dots, \uv_{k}$ such that 
$$
\delta_1(\ua, \un_1,\um_1)\wedge \delta_2(\ua,\un_2, \um_2)\wedge \bigwedge_{j=1}^k (\um_1\um_2\cap \uv_{j}\uguale \emptyset) 
$$
were not $T$-consistent. Putting $\uv$ equal to the tuple obtained by juxtaposition  
$\uv_{1} \cdots \uv_{k}$, we would  get a $\uv$ contradicting~\eqref{eq:consistent1}.

Let $\Theta_1$ be $\{\delta_1(\ua, \un_1,\um_1)\}\cup \{\um_1\cap \uv\uguale \emptyset\}_{\uv}$ and let 
$\Theta_2$ be $\{\delta_2(\ua, \un_2,\um_2)\}\cup \{\um_2\cap \uv\uguale \emptyset\}_{\uv}$. Since $\Theta_1\cup \Theta_2$ is 
equal to~\eqref{eq:consistent2} which is
$T$-consistent, there is no 
ground $\Sigma\cup\{\ua\}$-formula $\theta(\ua)$ such that $\Theta_1\vdash_T \theta(\ua)$ and such that $\Theta_2\cup\{\theta(\ua)\}$ is not $T$-consistent.
By Lemma~\ref{lem:app},
 we can then produce models $\cM_1, \cM_2$ of $T$ such that $\cM_1\models \Theta_1$, 
$\cM_2\models\Theta_2$ and such that the intersection of their supports is precisely the substructure generated by the interpretation of the constants $\ua$.
If we now strongly amalgamate them, we get a model of $T$ in which 
$\delta_1(\ua, \un_1,\um_1), \delta_2(\ua, \un_2,\um_2), \um_1\cap \um_2\uguale \emptyset$ are all true, showing that~$\eqref{eq:antecedent_var}$ fails.

\vskip 2mm
The implication 
${\rm (ii)}\Rightarrow {\rm (iii)}$ is trivial. We prove ${\rm (iii)}\Rightarrow {\rm (i)}$.
Suppose that we have~$\eqref{eq:antecedent}\Rightarrow\eqref{eq:consequent}$ and let us prove strong sub-amalgamability. If the latter property fails, by 
Robinson Diagram Lemma, there are models $\cM_1, \cM_2$ of $T$ together with a shared substructure $\cA$ such that the set of sentences 
 $$
\Gamma ~\coincide ~ \Delta_{\Sigma}(\cM_1)\cup \Delta_{\Sigma}(\cM_2)\cup \{ m_1\neq m_2~\vert~ m_1\in \vert \cM_1\vert\setminus \vert \cA\vert,
~ m_2\in \vert \cM_2\vert\setminus \vert \cA\vert\} 
$$ 
is not $T$-consistent. By compactness, the sentence
$$
\delta_1(\ua, \um_1)\wedge \delta_{2}(\ua, \um_2) \rightarrow \um_1\cap \um_2 \not\uguale \emptyset
$$
is $T$-valid, for some tuples $\ua\subseteq\vert \cA\vert$, $\um_1\subseteq(\vert \cM_1\vert\setminus \vert \cA\vert)$,
$\um_2\subseteq(\vert \cM_2\vert\setminus \vert \cA\vert)$ and for some ground formulae $\delta_1(\ua, \um_1),\delta_{2}(\ua, \um_2)$
true in $\cM_1, \cM_2$, respectively. By the implication~$\eqref{eq:antecedent}\Rightarrow\eqref{eq:consequent}$, there exists a finite tuple $\uv(\ua)$ 
of $\Sigma\cup\{\ua\}$-terms such that
$$
\delta_1(\ua, \um_1)\wedge (\um_1\cap \uv(\ua) \uguale \emptyset)\wedge  
\delta_{2}(\ua, \um_2) \wedge (\um_2\cap \uv(\ua) \uguale \emptyset)
$$
is not $T$-consistent. Since $T$ has quantifier-free interpolation, there is a ground formula $\theta(\ua)$ such that 
\begin{equation}
\label{eq:1}
\delta_1(\ua, \um_1)\wedge (\um_1\cap \uv(\ua) \uguale \emptyset)\to \theta(\ua) 
\end{equation}
is $T$-valid and 
\begin{equation}
\label{eq:2}
\delta_{2}(\ua, \um_2) \wedge (\um_2\cap \uv(\ua) \uguale \emptyset) \wedge \theta(\ua) 
\end{equation}  
is not $T$-consistent. 
However this is a contradiction: since $\um_1\subseteq \vert \cM_1\vert\setminus \vert \cA\vert$, the formula 
$\um_1\cap \uv(\ua) \uguale \emptyset$ is true in $\cM_1$, which 
 entails that $\theta(\ua)$ is true in $\cA$ and in $\cM_2$ too, where 
~\eqref{eq:2} consequently holds.
\myqed
\end{proof}

Notice that (iii) is just the special case of (ii) arising when the tuple $\uz$ is empty; 
this special case can be enough in the applications (for instance, the combined interpolation algorithm from Section~\ref{sec:algorithms} makes use of this special case only).

We now come to the results concerning equality interpolation and quantifier elimination.

\begin{lemma}
  \label{lem:qe} 
  Let $T$ be a theory admitting quantifier elimination; $T$ is
  universal iff for every quantifier-free formula $\phi (\ux, \uy)$,
  there exists tuples  $\ut_1(\ux), \dots, \ut_n(\ux)$ of tuples of terms such
  that
  \begin{equation}
    \label{eq:elimination}
    T\vdash \exists \uy \,\phi(\ux, \uy) \leftrightarrow \bigvee_{i=1}^n \phi(\ux, t_i(\ux)) ~~.
  \end{equation}
\end{lemma}
\begin{proof}
  If the condition of the Lemma is true for every $\phi(\ux, \uy)$,
  one can find an equivalent universal set of axioms for $T$ as
  follows.  Notice that the right-to-left side
  of~\eqref{eq:elimination} is a logical validity and the
  left-to-right side is equivalent to a universal formula. Thus, we
  can take as axioms for $T$ the universal closures of the
  left-to-right sides of~\eqref{eq:elimination}, together with the
  ground formulae which are logical consequences of $T$. In fact,
  axioms~\eqref{eq:elimination} are sufficient to find for every
  sentence a ground formula $T$-equivalent to it.
 
  Conversely, suppose that $T$ is universal and that there is
  $\phi(\ux, y_1, \dots, y_m)$ such that~\eqref{eq:elimination} does
  not hold (for all possible tuples of $m$-tuples of terms).  Then, by
  compactness, we have that the set of sentences
  $$
  \Gamma~\coincide~\{\phi(\ua, \ub)\}\cup \{\neg\phi(\ua, \ut(\ua))\}_{\ut}
  $$ 
  is $T$-consistent (here $\Sigma$ is the signature of $T$, $\ua,
  \ub:=b_1, \dots, b_m$ are tuples of fresh constants and $\ut$ vary
  on the set of $m$-tuples of $\Sigma\cup \{\ua\}$-terms). Let $\cM$
  be a $T$-model of $\Gamma$ and let $\cN$ be the substructure of
  $\cM$ generated by the $\ua$. Since $T$ is universal and truth of
  universal sentences is preserved under taking substructures, $\cN$
  is also a model of $T$ and since $T$ has quantifier-elimination,
  $\exists \uy \,\phi(\ua, \uy)$ - being $T$-equivalent to a
  quantifier-free $\Sigma\cup \{\ua\}$-sentence - is true in $\cN$
  too. This is a contradiction because from $\cN\models \exists \uy
  \,\phi(\ua, \uy)$ it follows that $\cN\models \phi(\ua, \ut(\ua))$ holds
  for some $\ut$, contrary to the fact that $\cM\not \models \phi(\ua,
  \ut(\ua))$ and to the fact that $\cN$ is a substructure of $\cM$.
\myqed
\end{proof}

\vskip 2mm\noindent
\textbf{Theorem}~\ref{thm:qe_strong_amalgamability}\emph{ 
  A universal theory admitting quantifier elimination is equality
  interpolating.
}
\begin{proof}
  We show that a universal and quantifier eliminable theory $T$
  satisfies the implication
  $\eqref{eq:antecedent}\Rightarrow\eqref{eq:consequent}$.  Suppose
  that $\eqref{eq:antecedent}$ holds; by the previous Lemma, there
  exists tuples of terms $\ut_1(\ux), \dots, \ut_k(\ux)$ such that
\begin{equation}
\label{eq:3}
\exists \uy_2 \,\delta_2(\ux, \uy_2) \leftrightarrow \bigvee_{j=1}^k \delta_2(\ux, \ut_j(\ux))
\end{equation}
is $T$-valid. For every $j=1, \dots, k$, if we replace $\uy_2$ with $\ut_j$ in $\eqref{eq:antecedent}$, we get
$$
\delta_1(\ux, \uy_1)\wedge \delta_2(\ux, \ut_j)\vdash_T \uy_1\cap \ut_j\neq \emptyset
$$
hence also
$$
\delta_1(\ux, \uy_1)\wedge \bigvee_{j=1}^k\delta_2(\ux, \ut_j)\vdash_T \bigvee_{j=1}^k (\uy_1\cap \ut_j\neq \emptyset)~~.
$$
Taking into account~\eqref{eq:3} and letting $\uv$ be the tuple $\ut_1\cdots\ut_k$ obtained by juxtaposition, we get
$$
\delta_1(\ux, \uy_1)\wedge \exists \uy_2\delta_2(\ux, \uy_2)\vdash_T  \uy_1\cap \uv\neq \emptyset~~.
$$
Removing the existential quantifier in the antecedent of the implication, we obtain
$$
\delta_1(\ux, \uy_1)\wedge \delta_2(\ux, \uy_2)\vdash_T  \uy_1\cap \uv\neq \emptyset
$$
and a fortiori $\eqref{eq:consequent}$, as desired.
\myqed
\end{proof}

We point out that the obvious converse of Theorem~\ref{thm:qe_strong_amalgamability} is not true: the theory of dense linear orders without endpoints has quantifier elimination, is equality interpolating (because it 
can be checked it has the strong 
sub-amalgamation property), but does not admit a universal set of axioms (because it is not closed under substructures).

The proof of {Theorem}~\ref{thm:qe_strong_amalgamability} is important also from the applications point of view. In fact, in the combined interpolation algorithm designed in Section~\ref{sec:algorithms},
one is given \formulae $\delta_1, \delta_2$ satisfying $\eqref{eq:antecedent}$ and is asked to compute terms $\uv(\ux)$ satisfying $\eqref{eq:consequent}$. In case our equality interpolating theory is universal 
and has quantifier elimination, one way to do this is to run the quantifier elimination algorithm over $\exists \uy_2 \,\delta_2(\ux, \uy_2)$ and to let $\uv$ be the tuple 
$\ut_1 \cdots \ut_k$
obtained by juxtaposition from the tuples in the right member of~\eqref{eq:3}.

Lemma~\ref{lem:qe} is also interesting in itself. According to Theorem~\ref{thm:qe_strong_amalgamability}, a sufficient condition for a theory $T$ to be equality interpolating is to have quantifier elimination 
via a universal set of axioms. The Lemma gives the possibility of checking the existence of such a set of axioms just by inspecting the quantifier elimination algorithm. Sometimes, this procedure is  easy. As an example,
 we can take the case of linear real arithmetic and Fourier-Motzkin algorithm. It is not difficult to see that Fourier-Motzkin algorithm satisfies the condition of Lemma~\ref{lem:qe} in the sense that it always `eliminates
existential quantifiers via tuples of terms'. For instance, when eliminating $\exists x$
from $\exists x\, ( x<y_1 \wedge x< y_2 \wedge y_3 < x)$ one gets
$$
(t_1<y_1 \wedge t_1< y_2 \wedge y_3 < t_1) \vee (t_2<y_1 \wedge t_2< y_2 \wedge y_3 < t_2) 
$$
where $t_1:=y_3+(y_1-y_3)/2$ and $t_2:=y_3+(y_2-y_3)/2$.

We now show that in the convex case, our notion of an equality interpolating theory coincides with the one given in~\cite{ym}.

\vskip 2mm\noindent
\textbf{Proposition}~\ref{prop:ym-eq-interpol}\emph{ 
  A convex theory $T$ having quantifier-free interpolation is equality
  interpolating iff for every pair $y_1, y_2$ of variables and for
  every pair of 
conjunctions of literals
$\delta_1(\ux,\uz_1,
  y_1), \delta_2(\ux,\uz_2,y_2)$ such that
  \begin{center} 
  \begin{tabular*}{\textwidth}{@{\extracolsep{\fill}}lcr}
    & 
    $\delta_1(\ux, \uz_1,y_1)\wedge \delta_2(\ux,\uz_2, y_2)\vdash_T y_1= y_2$ 
    & $(\ref{eq:ym_ant})$
  \end{tabular*}
  \end{center} 
  there exists a term $v(\ux)$ such that
  \begin{center} 
  \begin{tabular*}{\textwidth}{@{\extracolsep{\fill}}lcr}
    &
    $\delta_1(\ux, \uz_1, y_1)\wedge \delta_2(\ux, \uz_2, y_2)\vdash_T y_1= v  \wedge  y_2= v.$
    & $(\ref{eq:ym_cons})$
  \end{tabular*}
  \end{center} 
}
\begin{proof}
  If $\delta_1(\ux, \uz_1, y_1)\wedge \delta_2(\ux, \uz_2,
  y_2)\vdash_T y_1 = y_2$ holds and $T$ is equality interpolating,
 it
  follows that there are terms $\uv(\ux):=v_1(\ux), \dots, v_n(\ux)$
  such that
  \begin{equation}
    \label{eq:intermediate}
    \delta_1(\ux,\uz_1, y_1)\wedge
    \delta_2(\ux,\uz_2, y_2)\vdash_T
    \bigvee_{i=1}^n (y_1= v_i) \vee \bigvee_{i=1}^n (y_2= v_i).
  \end{equation}
  Let $w_1, \dots, w_n$ be fresh variables;
  from~\eqref{eq:intermediate} it follows that
  \begin{equation*}
    \delta_1(\ux,\uz_1, y_1)\wedge
    \delta_2(\ux,\uz_2, y_2)\wedge
    \bigwedge_{i=1}^n (w_i= v_i)\vdash_T 
    \bigvee_{i=1}^n (y_1= w_i) \vee \bigvee_{i=1}^n (y_2= w_i).
  \end{equation*}
  Applying convexity, we obtain that there is some $i$ such that
  either
  \begin{equation*}
    \delta_1(\ux,\uz_1, y_1)\wedge
    \delta_2(\ux,\uz_2, y_2)\wedge
    \bigwedge_{i=1}^n (w_i= v_i)\vdash_T 
    y_1= w_i
  \end{equation*}
  or
  \begin{equation*}
    \delta_1(\ux,\uz_1, y_1)\wedge
    \delta_2(\ux,\uz_2, y_2)\wedge 
    \bigwedge_{i=1}^n (w_i= v_i)\vdash_T 
    y_2= w_i
  \end{equation*}
  holds.  Replacing the $w$'s with the $v$'s, this gives either
  \begin{equation*}
    \delta_1(\ux,\uz_1, y_1)\wedge \delta_2(\ux,\uz_2, y_2)\vdash_T y_1= v_i
  \end{equation*}
  or
  \begin{equation*}
    \delta_1(\ux,\uz_1, y_1)\wedge \delta_2(\ux,\uz_2, y_2)\vdash_T y_2= v_i.
  \end{equation*}
  In both cases (taking into consideration~\eqref{eq:ym_ant}), we get
  $\delta_1(\ux,\uz_1, y_1)\wedge \delta_2(\ux,\uz_2, y_2)\vdash_T
  y_1= v_i \wedge y_2= v_i$, as required by~\eqref{eq:ym_cons}.

  Vice versa, when assuming the
  implication~$\eqref{eq:ym_ant}\Rightarrow \eqref{eq:ym_cons}$, it is
  very easy to show (by applying convexity) that $T$ is equality interpolating.\footnote{ 
Notice that in Definition~\ref{def:eq-interpol}, we can restrict $\delta_1, \delta_2$ to be conjunctions of 
literals,  getting anyway an equivalent definition. In fact, 
if~\eqref{eq:antecedent_var} holds,
 $\delta_1\equiv \bigvee_j \theta_{1j}$ and 
$\delta_2\equiv\bigvee_k\theta_{2k}$, 
then we can find tuples $\uv_{jk}$ satisfying $\theta_{1j}\wedge \theta_{2k}\vdash_T \uy_1\uy_2\cap \uv_{jk}\neq\emptyset$
and finally get by juxtaposition a tuple $\uv$ satisfying~\eqref{eq:consequent_var}. 
}
\myqed
\end{proof}

\subsection{Proofs for Section~\ref{sec:algorithms}}

In this Subsection we prove the relevant properties (soundness, completeness, termination) of our combined interpolation algorithm $\mathsf{CI}(T_1, T_2)$, where $T_1, T_2$ are two signature-disjoint, stably infinite and
equality interpolating theories whose SMT problems are decidable.

\begin{lemma}\label{lem:unsat}
 If rules \textbf{Decide}$_X$, \textbf{Share}$_i$ and
  \textbf{Terminate}$_i$ do not apply to a pair $A, B$, then $A\cup B$ is $T_1\cup T_2$-satisfiable, unless $\bot \in A\cup B$.
\end{lemma}

\begin{proof} Let $\ua, \uc$ the free constants occurring in $A$ and $\ub, \uc$ be the free constants occurring in $B$.
 If the above rules do not apply and $\bot\not\in A\cup B$, then 
  $A_i\cup B_i \cup
  \{\ua\cap \ub=\emptyset\}$ is $T_i$-satisfiable for $i=1, 2$; moreover $A$  contains an $A$-assignment $\alpha$ and $B$ contains a $B$-assignement $\beta$.
This means that $A_1\cup A_2$ entails $A$ and $B_1\cup B_2$ entails $B$, so that it is sufficient to show the $T_1\cup T_2$-satisfiability of $A_1\cup A_2\cup B_1\cup B_2$ only. 
The latter follows from Lemma~\ref{lem:NO}, because the sets  
$$
\Theta_i \equiv A_i\cup B_i \cup
  \{\ua\cap \ub=\emptyset\}
$$
 satisfy the hypothesis of the Lemma. Pick in fact a pair of constants $d_1, d_2$ from $\ua, \ub, \uc$: if they are both from $\ua,\uc$ or both from $\ub, \uc$, then either $d_1=d_2$ or 
$d_1\neq d_2$ belongs to $\Theta_i$, as $\alpha\cup \beta$ has assigned a truth value to $d_1=d_2$. If one of them is in $\ua$ and the other is in $\ub$, then $d_1\neq d_2\in \Theta_i$
by construction.
\myqed
\end{proof}

\vskip 2mm\noindent
\textbf{Theorem}~\ref{thm:main} \emph{
  Let $T_1$ and $T_2$ be two signature disjoint, stably-infinite, and
  equality interpolating theories having decidable SMT problems.
  Then, $\mathsf{CI}(T_1,T_2)$ is a quantifier-free interpolation
  algorithm for the combined theory $T_1\cup T_2$.
}

\vskip 2mm
\begin{proof} 
Let $A_0, B_0$ be our input $T_1\cup T_2$-unsatisfiable pair. By repeatedly applying our transformations \textbf{Decide}$_X$, \textbf{Share}$_i$ and
  \textbf{Terminate}$_i$ to it, we produce a tree $\tau$ (the pairs labeling the successors of a node are the possible outcomes of our transformations, which are non deterministic).
Clearly \textbf{Decide}$_X$, \textbf{Share}$_i$ and
  \textbf{Terminate}$_i$ are satisfiability-preserving, in the sense that a pair to which they are applied is $T_1\cup T_2$-satisfiable iff one of the outcomes is. As a consequence, by Lemma~\ref{lem:unsat},
$\bot$ must belongs to all pairs labeling the leaves. Thus, since \textbf{Decide}$_X$, \textbf{Share}$_i$ and
  \textbf{Terminate}$_i$ can all be justified by metarules, our tree $\tau$ is an interpolating metarules refutation (and we are done by Proposition~\ref{prop:metarules}),  \emph{provided we show that 
$\tau$ is finite}. Finiteness of $\tau$ is also needed to prove the termination of our algorithm.

We apply K\"onig Lemma and show that all branches of $\tau$ are finite. Notice that the transformation \textbf{Decide}$_X$ can be applied many times in a branch: this is because \textbf{Share}$_i$ introduces a new
ground formula $\theta$ and alters the definition of an $A$-relevant and a $B$-relevant atom (it introduces new \abcommon constants by Term Sharing). However, \textbf{Share}$_i$ can be applied only
finitely many times, as it decreases the number of \astrict or \bstrict constants. Once \textbf{Share}$_i$ is no more applied, just single applications of \textbf{Decide}$_A$, \textbf{Decide}$_B$, 
\textbf{Terminate}$_i$ are possible.
 \myqed
\end{proof}

\newpage 

\section{Quantifier elimination through universal axioms}
\label{app:qe}

In this Appendix we give details concerning a couple of applications of Theorem~\ref{thm:qe_strong_amalgamability}.

\subsection{Integer Linear Arithmetic}
\label{subapp:lai}

Presburger Arithmetic \PrA is the theory so specified. Its signature consists of the symbols $0,1, +, -, <$ in addition to the infinite predicates $P_n$ (one for every $n>0$).
A set of axioms for \PrA is the following one
\begin{eqnarray*}
  \begin{array}{l}
 \forall x,y,z.~ x+(y+z)= (x+y)+z \quad
  \\
  \forall x,y.~ x+y= y+x \quad
  \\
  \forall x.~x+0=x \quad
  \\
  \forall x.~x+(-x)=0 \quad
\\ 
  \forall x. ~  x\not<x \quad
 \\
 \forall x,y,z.~(x<y \wedge y<z \to x<z) \quad
 \\
 \forall x,y.~ x<y \vee x=y \vee y<x
\quad 
\\
 \forall x,y,z.~x<y\to x+z<y+z, \quad 
  \\
   0<1, \quad 
  \\
  \forall y.\neg(0<y\wedge y<1),
  \quad 
  \\
 \forall x\, \exists y.~ \bigvee_{0\leq r<n} x=ny+r \quad
 \\
 \forall x.~ P_n(x) \leftrightarrow \exists y\;(ny=x)\quad 
  \end{array}
\end{eqnarray*}
(we used the abbreviations $nt$ for the sum of $n$-copies of $t$ and $n$ for $n1$).
Presburger arithmetic enjoys quantifier-elimination: a detailed proof can be found e.g. in~\cite{enderton} or also in the online available notes\footnote{\url{http://www.math.uiuc.edu/~vddries/}}
 L. Van Der Dries ``Mathematical Logic Lecture Notes''
(where we took the above axiomatization from). However, \PrA is not equality interpolating because $\PrA\cup \EUF$ does not enjoys quantifier-free interpolation~\cite{BKR+2010}.

In Subsection~\ref{subsec:eq-interpol-at-work}, we proposed the theory \LIA, comprising in its language also  the unary function symbols
$\mydiv{}{n}$ (representing integer division by $n$, for $n>1$). In \LIA, one can define $P_n(x)$ as $\myrem{x}{n}=0$  (recall that $\myrem{x}{n}$ abbreviate $x-n(\mydiv{x}{n})$).
Using this definition, we can view \LIA as a supertheory of \PrA, because all the axioms of \PrA are derivable in \LIA.\footnote{For the last one, 
show that the following universal sentences are derivable in \PrA for every $n>0$:
$$
\forall x.~nx=0\to x=0 \qquad \forall x.~\bigwedge_{0<r<n} nx\neq r~.
$$
} 
 We are ready to show that Theorem~\ref{thm:qe_strong_amalgamability} applies to \LIA:
\begin{proposition}
 \LIA is equality interpolating.
\end{proposition}

\begin{proof}
 In view of Theorem~\ref{thm:qe_strong_amalgamability}, since \LIA is universal,
 we only need to show that \LIA has elimination of quantifier.
 Let $\phi(\ux)$ be an abritrary formula of \LIA; consider an atom $L$ occurring in $\phi$ containing an occurrence of a term $u$ of the kind $\mydiv{t}{n}$. 
Modulo \LIA, the atom $L$ is equivalent to
\begin{eqnarray}\label{eq:replace}
\exists y\; \bigvee_{0\leq r<n} (t=ny+r  \wedge L[y/u])
\end{eqnarray}
(this is because $\bigvee_{0\leq r<n} (t=ny+r)\leftrightarrow y=\mydiv{t}{n}$ follows from the axioms of \LIA).
We can then replace $L$ by~\eqref{eq:replace} in $\phi$ and get an equivalent formula. If we do this  exhaustively, we obtain a formula $\phi'$ such that 
$\LIA\vdash \phi \leftrightarrow \phi'$. Since, as we observed above, \LIA is a supertheory of \PrA and the latter enjoys quantifier elimination, we can find a quantifier-free $\phi''(\ux)$
 such that $\LIA\vdash \phi \leftrightarrow \phi''$.
\myqed
\end{proof}

\subsection{Unit-Two-Variable-Per-Inequality}
\label{subapp:utvpi}

This theory (called \UTVPI in the literature) is another interesting fragment of integer linear arithmetic, slightly more expressive than \IDL. If can be defined as the theory 
whose axioms are the sentences true in $\mathbb{Z}$ in the signature comprising predecessor $pred$, successor $succ$, $0,<$ and $-$ (the latter is viewed as a unary symbol). We shall exhibit here a  set of 
universal quantifier eliminating axioms for 
 \UTVPI (thus showing that \UTVPI is equality interpolating too, thanks to Theorem~\ref{thm:qe_strong_amalgamability}).

Like in the case of \IDL, let us examine the shape of the atoms of \UTVPI. They  are equivalent to formulae having the form $\pm i \bowtie
f^n(j)$ (for $n\in \mathbb{Z}$, $\bowtie\,\in \{=, <, >\}$)\footnote{
We use $>$ as a defined symbol ($t>u$ stands for $u<t$).
} where $i,j$ are variables or the
constant $0$, $f^0(j)$ is $j$, $f^{k}(j)$ abbreviates
$succ(succ^{k-1}(j))$ when $k>0$ or $pred(pred^{k-1}(j))$ when $k<0$.
(Usually, $\pm i \bowtie f^n(j)$ is written as $i \pm j\bowtie n$ or as $i \bowtie n \pm j$).  

\begin{proposition}
 \UTVPI is equality interpolating.
\end{proposition}

\begin{proof}
 We take inspiration from Lemma~\ref{lem:qe}, that is we directly supply a quantifier elimination algorithm for \UTVPI
satisfying~\eqref{eq:elimination} (thus, the left to right sides of \formulae~\eqref{eq:elimination} will be the relevant axiomatization for \UTVPI, once joined with the 
universal sentences in the signature of \UTVPI which are true in $\mathbb{Z}$).\footnote{
The latter are needed to normalize all atoms to the form $i \bowtie n \pm j$.
} 
As usual, it is sufficient to eliminate single existentially quantified variables from primitive \formulae~\cite{CK}.
This means that, since negation can be eliminated, we must consider formulae $\exists x\,\phi$  where $\phi$ is a conjunction of atoms of the following kinds:
$$
x= m_i \pm t_i, \qquad x < m_j\pm u_j, \qquad  x > m_k \pm v_k,
$$ 
where $x$ does not occur in the $t_i,u_j, v_k$ (otherwise either $\phi$ is inconsistent or the atom is redundant or it simplifies to an atom of the above kinds).
If there are literals of the first kind, the quantifier $\exists x$ can be eliminated by substitution (this schema fits~\eqref{eq:elimination}), so suppose there are none.
If there are no literals of the second kind or no literals of the third kind, $\exists x\, \phi$ is equivalent to $\top$ (use 
the terms $pred(m_j\pm u_j), succ(m_k\pm v_k)$ to fit~\eqref{eq:elimination}).
If there are both literals of the second and of the third kind, $\exists x\,\phi$ is equivalent to $\bigvee_k \phi(succ(m_k \pm v_k))$.
\myqed
\end{proof}

\newpage

\section{Equality interpolating and Beth definability}\label{app:beth}

In this Section we discuss the connection of the notions introduced in this paper with standard topics in mathematical logic and universal algebra.
This complementary material is included here for the sake of completeness.

\emph{Beth definability theorem}~\cite{CK} is a classical  result in model theory; we show that in the convex case equality 
interpolating can be interpreted as a `modulo theory'
version of a Beth definability property. We find in the non-convex case too a `Beth-like' formulation of equality interpolation. 
In the end, we use the Beth definability formulation of equality interpolation in order to briefly discuss 
the relationship between  our results and well-known results concerning strong amalgamation from the algebraic literature.

To begin with, we add a further equivalent characterization to the list (i)-(iii) 
 of Theorem~\ref{thm:strong_amalgamation_syntactic}:

\vskip 2mm\noindent
\textbf{Theorem}~\ref{thm:strong_amalgamation_syntactic}\emph{
The following conditions are equivalent for a theory $T$ having quan\-ti\-fier-free interpolation:
\begin{description}
 \item[{\rm (i)}] $T$ is strongly sub-amalgamable;
 \item[{\rm (ii)}] $T$ is equality interpolating;
 \item[{\rm (iii)}] $T$ satisfies the implication $\eqref{eq:antecedent}\Rightarrow \eqref{eq:consequent}$ (for every $\delta_1, \delta_2$);
\item[{\rm (iv)}]
for every quantifier-free formula $\delta(\ux, \uz, \uy)$ such that
\begin{equation}\label{eq:beth_gen_ant}
 \delta(\ux, \uz', \uy') \wedge \delta(\ux, \uz'', \uy'')\vdash_T \uy'\cap \uy'' \not= \emptyset
\end{equation}
there are terms $\uv(\ux)$ such that
\begin{equation}\label{eq:beth_gen_cons}
 \delta(\ux, \uz, \uy) \vdash_T \uy\cap \uv\not=\emptyset.
\end{equation}
\end{description}
}

\begin{proof}
 We already proved (in the previous formulation of Theorem~\ref{thm:strong_amalgamation_syntactic} in Appendix~\ref{app:proofs}) that conditions (i)-(ii)-(iii) are 
all equivalent to each other.

Assume (iv) and~$\eqref{eq:antecedent}$. Take $\uy:=\uy_1, \uy_2$ and put $\delta(\ux, \uy):=\delta_1(\ux, \uy_1)\wedge \delta_2(\ux, \uy_2)$. 
Now notice that 
$\delta(\ux, \uy_1', \uy_2') \wedge \delta(\ux, \uy_1'', \uy_2'')$ is 
$$
\delta_1(\ux, \uy'_1)\wedge \delta_2(\ux, \uy'_2)\wedge \delta_1(\ux, \uy''_1)\wedge \delta_2(\ux, \uy''_2);
$$
since by~$\eqref{eq:antecedent}$ we have 
$$
\delta_1(\ux, \uy'_1)\wedge \delta_2(\ux, \uy''_2)\vdash_T \uy'_1\cap \uy''_2\neq \emptyset
$$ 
a fortiori we get
\begin{equation}
 \delta(\ux, \uy_1', \uy_2') \wedge \delta(\ux, \uy_1'', \uy_2'')\vdash_T \uy_1'\uy_2'\cap \uy_1''\uy_2'' \not= \emptyset,
\end{equation}
By (iv),
there are terms $\uv(\ux)$ such that $\delta(\ux, \uy_1, \uy_2)\vdash_T \uy_1\uy_2\cap \uv \not= \emptyset$, which is the same as~\eqref{eq:consequent}.

For the vice versa, we suppose that~$\eqref{eq:antecedent_var}~\Rightarrow~\eqref{eq:consequent_var}$ holds. Consider $\delta(\ux,\uz, \uy)$ such that~\eqref{eq:beth_gen_ant} holds.
Then, we can find $\uv(\ux)$ such that 
\begin{equation}
 \delta(\ux, \uz', \uy') \wedge \delta(\ux, \uz'', \uy'')\vdash_T (\uy'\cap \uv \not= \emptyset) \vee (\uy''\cap \uv \not= \emptyset)
\end{equation}
holds. Making the substitutions $\uz'\mapsto \uz, \uz''\mapsto \uz, \uy'\mapsto \uy, \uy''\mapsto \uy$, this gives precisely~\eqref{eq:beth_gen_cons}. 
\myqed
\end{proof}

Condition (iv) above can be interpreted as a `generalized Beth property'. The situation becomes clearer in the simplified convex case; we first
restate Proposition~\ref{prop:ym-eq-interpol}:

\vskip 2mm\noindent
\textbf{Proposition}~\ref{prop:ym-eq-interpol}\emph{ 
The following conditions are equivalent for  a
  convex theory $T$ having quantifier-free interpolation:
\begin{description}
\item[{\rm (i)}] $T$ is equality interpolating;
 \item[{\rm (ii)}] $T$ satisfies the implication $\eqref{eq:ym_ant}\Rightarrow \eqref{eq:ym_cons}$ (for every conjunctions of literals $\delta_1, \delta_2$);
 \item[{\rm (iii)}] for every pair $\ux, \uz$ of tuples of variables, for every further variable $y$  and for
 every conjunction of literals
$\delta(\ux, \uz, y)$ such that
\begin{equation*}
 \delta(\ux, \uz', y') \wedge \delta(\ux, \uz'', y'')\vdash_T y' = y''~,
\end{equation*}
there is a  term $v(\ux)$ such that
\begin{equation*}
\delta(\ux, \uz, y) \vdash_T y = v~.
\end{equation*}
\end{description}
}

\begin{proof} 
Again, we already know from Appendix~\ref{app:proofs} that (i) and (ii) are equivalent.

Assume that
${\rm (iii)}$ holds and consider $\delta_1(\ux, \uz_1, y_1), \delta_2(\ux, \uz_2, y_2)$ satisfying $\eqref{eq:ym_ant}$.
Take  $\delta(\ux, \uz_1, \uz_2,y):=\delta_1(\ux,\uz_1,y)\wedge \delta_2(\ux,\uz_2, y)$. Now 
$\delta(\ux, \uz'_1, \uz'_2,y')\wedge \delta(\ux, \uz''_1, \uz''_2,y'')$ is 
$$
\delta_1(\ux,\uz'_1,y')\wedge \delta_2(\ux,\uz'_2, y') \wedge \delta_1(\ux,\uz''_1,y'')\wedge \delta_2(\ux,\uz''_2, y''),
$$
hence (considering the first and the fourth conjunct) from $\eqref{eq:ym_ant}$ we get 
$$
\delta(\ux, \uz'_1, \uz'_2,y')\wedge \delta(\ux, \uz''_1, \uz''_2,y'')\vdash_T y'=y''.
$$
By ${\rm (iii)}$, there is a term $v(\ux)$ such that
\begin{equation}\label{eq:bth}
\delta_1(\ux,\uz_1,y)\wedge \delta_2(\ux,\uz_2, y)\vdash_T y=v(\ux). 
\end{equation}
Again by $\eqref{eq:ym_ant}$, we obtain (after renamings)
$$
\delta_1(\ux,\uz_1,y_1)\wedge \delta_2(\ux,\uz_2, y_2)\vdash_T y_1=y_2\wedge \delta_2(\ux,\uz_2, y_1) ~;
$$
thus (taking into account~\eqref{eq:bth}) also 
$$
\delta_1(\ux,\uz_1,y_1)\wedge \delta_2(\ux,\uz_2, y_2)\vdash_T y_1=y_2\wedge y_1=v(\ux)
$$
and finally $\eqref{eq:ym_cons}$.

Vice versa, if (ii) holds and we have $\delta(\ux, \uz', y') \wedge \delta(\ux, \uz'', y'')\vdash_T y' = y''$, we can find 
$v(\ux)$ such that 
$$
\delta(\ux,\uz',y')\wedge \delta(\ux,\uz'', y'')\vdash_T y'=v\wedge y''=v~;
$$
applying the substitution 
$\uz'\mapsto \uz, \uz''\mapsto \uz, \uy'\mapsto \uy, \uy''\mapsto \uy$, this gives our claim $\delta(\ux, \uz, y) \vdash_T y = v$.
\myqed
\end{proof}

A \emph{primitive}
formula is obtained from a 
conjunction of literals
by prefixing to it a string of existential quantifiers. 
We can reformulate the condition (iii) from Proposition~\ref{prop:ym-eq-interpol} above as follows:
 \begin{description}
 \item[{\rm (iii)'}]
 for every tuple of variables $\ux$, for every further variable $y$  and for every 
\emph{primitive}
 formula $\theta(\ux, y)$
 such that
 $\theta(\ux, y') \wedge \theta(\ux, y'')\vdash_T y' = y'' $,
there is a  term $v(\ux)$ such that
$ \theta(\ux, y) \vdash_T y = v.$
\end{description}
This is precisely \emph{Beth definability property}~\cite{CK}, modulo $T$, for 
primitive
\formulae.
 Hence equality interpolating coincides with this 
`primitive
Beth definability property' in the convex case.

To conclude, for the interested reader, we make some observations connecting the above result with the algebraically oriented literature (see~\cite{tholen}
for a survey and for pointers to relevant papers).
In an appropriate context from universal algebra, strong amalgamability is shown  to be equivalent to the conjunction of amalgamability and
of regularity of epimorphisms
(alternatively: and of regularity of monomorphisms). In the same context, unravelling the definitions and using presentations of algebras as quotient of free ones, 
it is not difficult to realize that the 
primitive
Beth definability property above is equivalent to regularity of monomorphisms. Thus,
 \emph{our results perfectly match with the algebraic characterization of strong amalgamability}. Our approach, however, is \emph{orthogonal} to  algebraic and 
category-theoretic approaches: such approaches are able in fact to prove characterizations of strong amalgamability that work in abstract sufficiently 
complete/cocomplete
categories, including consequently categories having nothing to do with  models of first order theories. On the other hand, existence of
minimal categorical structure 
fails in our context as soon as we go beyond the universal Horn case. Thus, the two approaches are incomparable and this is reflected by the different 
techniques employed (we mostly rely on diagrams and compactness, whereas the category-theoretic approach mostly exploit universal properties).

\newpage

\section{Interpolation with free fuction symbols}\label{app:general}

In this paper, we  treated quantifier free interpolation only with respect to variables, in the sense that we always considered all non variable symbols as shared symbols.
This is not the notion of interpolation commonly used in verification, where also free functions and predicate symbols are not allowed to apper in the interpolants in case they do not occur in 
both the \formulae to be interpolated. We show here  that this more general notion of quantifier free 
interpolation can be reduced to combined interpolation  and thus that it is equivalent to strong sub-amalgamability too.  

\begin{definition}\label{def:general}
Let $T$ be a theory in a signature $\Sigma$; we say that $T$ has the \emph{general quantifier-free interpolation property} iff for every signature $\Sigma'$ (disjoint from $\Sigma$) 
and for every finite sets of ground $\Sigma\cup\Sigma'$-\formulae
$A, B$ such that $A\wedge B$ is $T$-unsatisfiable,\footnote{By this (and similar notions) we mean that $A\wedge B$ is unsatisfiable in all $\Sigma'$-structures whose $\Sigma$-reduct is a model of $T$. We use the same convention 
as in Section~\ref{sec:algorithms} and indicate with the letters $A, B$ both a finite set of ground \formulae and its conjunction.
} 
there is a ground formula $\theta$ such that: (i) $A$ $T$-entails $\theta$;  (ii) $\theta\wedge 
B$ is $T$-unsatisfiable; (iv) all predicate, constants and function symbols from $\Sigma'$ 
occurring in $\theta$ occur also in $A$ and in $B$. 
\end{definition}

Notice that the above definition becomes equivalent to the definition of quantifier free interpolation property introduced in Section~\ref{sec:background} 
if we restrict it to the  signatures
 $\Sigma'$ containing only constant function symbols. One may wonder whether 
Definition~\ref{def:general} is the same as asking for quantifier free interpolation for all combined theoris $T\cup \EUF(\Sigma')$;  at a first glance, it does not seem to be so because in Definition~\ref{def:general}
we require also that the function and the predicate symbols from $\Sigma'$ not occurring in both $A, B$ do not occur in $\theta$ either. We shall see however that such symbols are immaterial because  they can be removed.

Let us fix a theory $T$ in a signature $\Sigma$ and let $\Sigma'$ be a further signature (disjoint from $\Sigma$). A finite set $A$ of ground $\Sigma\cup \Sigma'$-\formulae is said to be \emph{$\Sigma_0$-flat}
(for some $\Sigma_0\subseteq \Sigma'$) iff $A$ is of the kind $A_0\cup A_1$, where $A_1$ does not contain $\Sigma_0$-symbols and $A_0$ is a set of literals of the kind
$$
f(a_1, \dots, a_n)= b, \quad P(a_1, \dots, a_n), \quad \neg P(a_1, \dots, a_n)
$$
where $f, P \in \Sigma_0$ and $a_1, \dots, a_n,b$ are constants not in $\Sigma_0$.

\begin{lemma}\label{lem:elimination}
 Let $T, \Sigma, \Sigma'$ be as above and let the finite set of ground $\Sigma\cup \Sigma'$-\formulae $A$ be $\Sigma_0$-flat (for some $\Sigma_0\subseteq \Sigma'$). Then it is possible to find a finite set of ground \formulae
 $A^{-\Sigma_0}$ such that: (i) $A^{-\Sigma_0}$ does not contain $\Sigma_0$-symbols; (ii) $A$ $T$-entails $A^{-\Sigma_0}$; (iii) $A^{-\Sigma_0}$ is $T$-satisfiable iff $A$ is $T$-satisfiable.
\end{lemma}

\begin{proof}
Let $A$ be $A_0\cup A_1$ as prescribed in the definition of $\Sigma_0$-flatness. We take as $A^{-\Sigma_0}$ the set of ground \formulae $A_0'\cup A_1$ where $A_0'$ is built as follows. For every function symbol $f\in \Sigma_0$
and for every pair of atoms $f(a_1, \dots, a_n)=b, f(a'_1, \dots, a'_n)=b'$ belonging to $A_0$ we include in $A'_0$ the  ground clause
\begin{equation}\label{eq:functions}
a_1=a'_1 \wedge \cdots \wedge a_n = a'_n \to b=b';
\end{equation}
similarly, for every predicate symbol $P\in \Sigma_0$  and for  every pair of literals $P(a_1, \dots, a_n)$, $\neg P(a'_1, \dots, a'_n)$ belonging to $A_0$ we include in $A'_0$ the  ground clause
\begin{equation}\label{eq:predicates}
a_1=a'_1 \wedge \cdots \wedge a_n = a'_n \to \bot.
\end{equation}
That $A'_0\cup A_1$ enjoys properties (i)-(ii) is clear; it remains to show that if it is $T$-satisfiable, so  is $A_0\cup A_1$.\footnote{The right-to-left side of (iii) is a consequence of (ii).}
 Suppose indeed that $\cM$ is a $\Sigma\cup (\Sigma'\setminus \Sigma_0)$-model of $T$ in which $A'_0\cup A_1$ is true. We expand $\cM$ to a $\Sigma\cup \Sigma'$-structure
 as follows. Let $f\in \Sigma_0$ have arity $n$ and let $c_1, \dots, c_n$ be elements from the support of $\cM$; 
then $f^{\cM}(c_1, \dots, c_n)$ is arbitrary, unless there are $f(a_1, \dots, a_n)=b\in A_0$ such that $c_1=a_1^{\cM}, \dots, c_n=a^{\cM}_n$: in this case, we put  $f^{\cM}(c_1, \dots, c_n)$ to be equal to $b^{\cM}$.
Since $\cM$ is a model of the clauses~\eqref{eq:functions}, the definition is correct. Similarly, if $P\in \Sigma_0$ has arity $n$, then $P^{\cM}$ is the set of $n$-tuples
 $c_1, \dots, c_n$ of elements from the support of $\cM$ such that there exists $P(a_1, \dots, a_n)\in A_0$ such that $c_1=a_1^{\cM}, \dots, c_n=a^{\cM}_n$. The literals from $A_0$ turns out to be all true by construction and 
because in $\cM$ the clauses~\eqref{eq:predicates} hold.
\end{proof}

\begin{theorem}
$T$ has the general quantifier free interpolation property iff it is strongly sub-amalgamable iff it is equality interpolating. 
\end{theorem}

\begin{proof}
 Since the general quantifier free interpolation property for $T$ implies the (ordinary) quantifier free interpolation property for all the theories $T\cup \EUF(\Sigma')$, 
it is clear from Theorem~\ref{prop:strong_amalgamation_needed} that the general quantifier free interpolation property implies strong sub-amal\-ga\-ma\-bi\-li\-ty.
 To show the vice versa, we use our metarules and Lemma~\ref{lem:elimination} above.

Let $\Sigma$ be the signature of $T$ and let $\Sigma'$ be disjoint from $\Sigma$; fix also finite sets of ground $\Sigma\cup \Sigma'$-formulae $A, B$
such that $A\wedge B$ is $T$-unsatisfiable. Let $\Sigma_A$ be the set of predicate and (non constant) function symbols from $\Sigma'$ that occur in $A$ but not in $B$; similarly, let $\Sigma_B$ be the set of 
predicate and (non constant) function symbols from $\Sigma'$ that occur in $B$ but not in $A$. We show how to transform $A$ into a $\Sigma_A$-flat $\tilde A$ by using  metarules (a similar transformation 
is applied to 
$B$ to get  a $\Sigma_B$-flat $\tilde B$). Using metarules (Define1), (Redplus1), (Redminus1) we can add `defining atoms' 
$f(a_1, \dots, a_n)=a$ (with fresh $a$) and replace all occurrences of the term $f(a_1, \dots, a_n)$ in $A$ by $a$; if we do it repeatedly, $A$ gets flattened, in the sense that function and 
predicate symbols (different from identity) in $A$ are always 
applied to constants. With the same technique, we can transform $A$ into a conjunction of defining atoms and ground \formulae in which function symbols from $\Sigma_A$ do not occur.
To take care of predicate symbols $P\in \Sigma_A$, we need guessings and metarule (Disjunction1): for every atom $P(a_1, \dots, a_n)$ occurring in $A$, we add 
either  $P(a_1, \dots, a_n)$ or  $\neg P(a_1, \dots, a_n)$ to $A$ and replace $P(a_1, \dots, a_n)$ with $\top$ or $\bot$, respectively 
(notice that because of such guessings the transformation from $A, B$ to $\tilde A, \tilde B$ may be non-deterministic).
 Since metarules are satisfiability-preserving and are endowed with recursive instructions for computation of interpolants, 
 it will be sufficient to find a desired interpolant $\theta$ for $\tilde A$ and $\tilde B$.

If we apply the tranformations of Lemma~\ref{lem:elimination}
to $\tilde A\cup \tilde B$ we can get $(\tilde A\cup \tilde B)^{-\Sigma_A}\coincide {\tilde A}^{-\Sigma_A}\cup \tilde B$ with the properties (i)-(iii) stated in that Lemma: in particular, function and predicate symbols   
from $\Sigma_A$ do not occur anymore in ${\tilde A}^{-\Sigma_A}$. We do the same for $\tilde B$ and eventually we get $\bar A, \bar B$ such that (a) $\bar A\cup \bar B$ is $T$-unsatisfiable; (b) $\tilde A$  $T$-entails $\bar A$,
$\tilde B$ $T$-entails $\bar B$; (c) all predicate and (non constant) functions symbols occurring in $\bar A$ occur also in $\bar B$ and vice versa. Let $\Sigma_C$ be the set of predicate and (non constant) function symbols occurring 
in  both $\bar A$ and $\bar B$. Since $T$ is strongly amalgamable, by Theorem~\ref{prop:strong_amalgamation_needed}, $T\cup \EUF(\Sigma_C)$ has the quantifier-free interpolation property.\footnote{The proof of the right-to-left side 
of that Theorem does not need the requirement that $\Sigma_C$ has at least a unary predicate symbol.}
Thus, there exists a ground formula $\theta$ containing, besides interpreted symbols from $\Sigma$, only predicate and function symbols from $\Sigma_C$, as well as individual free constants occurring both in $\bar A$ and in $\bar B$,
such that $\bar A$ $T$-entails $\theta$ and $\bar B\wedge \theta$ is $T$-inconsistent.  By (b) above, we get that $\tilde A$ $T$-entails $\theta$ and $\tilde B\wedge \theta$ is $T$-inconsistent, thus $\theta$ is the desired
interpolant.

The equivalence between strong sub-amalgamability and equality interpolating property comes from Theorem~\ref{thm:strong_amalgamation_syntactic}.
\end{proof}

\newpage

\section{A counterexample: golden cuff links }\label{app:counterexample}

Here we show by exhibiting a formal counterexample that 
the 
`convex' formulation of the equality interpolating property is not sufficient to guarantee the modularity of quantifier-free interpolation
for non-convex theories. Intuitively, the reason is that disjunctions of equalities must be propagated in the non convex case and the convex formulation of the equality interpolation property does not say anything about them.
This Appendix can be read independently on the remaining part of the Technical Report.

We say that a theory $T$ has the \emph{YMc property} (`convex Yorsh-Musuvathi property') iff it has quantifier-free interpolation property and moreover the implication
$\eqref{eq:ym_ant} \Rightarrow \eqref{eq:ym_cons}$ holds, i.e. 
for every pair $y_1, y_2$ of variables and for
  every pair of 
conjunctions of literals
$\delta_1(\ux,\uz_1,
  y_1), \delta_2(\ux,\uz_2,y_2)$ such that
  \begin{equation*}
    \delta_1(\ux, \uz_1,y_1)\wedge
    \delta_2(\ux,\uz_2, y_2)\vdash_T
    y_1= y_2
  \end{equation*}
  there exists a term $v(\ux)$ such that
  \begin{equation*}
    \delta_1(\ux, \uz_1, y_1)\wedge
    \delta_2(\ux, \uz_2, y_2)\vdash_T 
    y_1= v  \wedge  y_2= v.
  \end{equation*}

To build our counterexample, we introduce a theory $CL$ which is meant to describe a set of \emph{cuff links}, containing at most one pair of \emph{golden} cuff links.
Formally, in the signature $\Sigma_{CL}$ of $CL$ we have a unary function symbol $(-)'$ and a unary predicate $G$.\footnote{A free constant $c_0$ is added to the signature $\Sigma_{CL}$ to prevent it from being empty.} 
The axioms of $CL$ say that $(-)'$ denotes the `twin' cuff link
$$
\forall x.~ x''=x, \quad \forall x.~ x\neq x'
$$
that twin cuff links are both golden or not
$$
\forall x.~ G(x) \leftrightarrow G(x')
$$
and that there is at most one pair of golden cuff links:
\begin{equation}\label{eq:unique}
\forall x\forall y.~G(x) \wedge G(y)\to x=y \vee x'=y.
\end{equation}

\begin{lemma}
 $CL$ has the quantifier free interpolation property, because it has the sub-amalgamation (but not the strong sub-amalgamation) property.
\end{lemma}

\begin{proof}
 That the sub-amalgamation property holds is quite clear: suppose we are given models $\cM_1, \cM_2$ of $CL$ sharing the substructure $\cA$ (as a side remark, notice that $\cA$ is also a model of $CL$ because $CL$ is universal).
As usual, we assume that  the intersection of the supports of $\cM_1$ and $\cM_2$ is the support of $\cA$. To amalgamate $\cM_1, \cM_2$ over $\cA$, it is sufficient to take the union of the supports of $\cM_1$ and $\cM_2$, with just one proviso:
if $\cM_1, \cM_2$ both contain a pair of golden cuff links that is not from $\cA$, then such pairs must be merged (the need of such merging is precisely what shows that strong sub-amalgamation fails). 
\end{proof}

\begin{proposition}
 $CL$ has the YMc property.
\end{proposition}
 
\begin{proof}
 We shall work with free constants (instead of with variables).
 Consider  finite sets of ground 
literals $A$, $B$ in the signature $\Sigma_{CL}$
enriched with additional free constants (let $\Sigma_A$ be the signature of $A$ and $\Sigma_B$ be the signature of $B$). We call \abcommon the ground terms built up from free constants occurring both in $A$ and in $B$;
ground terms built up from constants occurring  in $A$ but not in $B$ are called \astrict (\bstrict ground terms are defined symmetrically). We call a ground term or literal \emph{pure} iff
 it is either from $\Sigma_A$ or from $\Sigma_B$. We argue by contraposition.
Suppose that, for an \astrict constant $a$ and a \bstrict constant $b$,  there is no \abcommon ground term $t$ such that $A\cup B\vdash_{CL} t=a\wedge t=b$; we show that $A\cup B\not\vdash_{CL} a=b$ by exhibiting a $\Sigma_A\cup\Sigma_B$-model $\cM$ of $CL$ such that $\cM\models A, \cM\models B$ and $\cM\not \models a=b$.

We can freely make further assumptions on our $A, B$: first, we can assume that there is at least 
one \abcommon ground term.\footnote{Because $\Sigma_{CL}$ has one.}, that terms like $d''$ do not occur in $A\cup B$, \footnote{
Because they simplify to $d$.} and that if a term occurs in $A\cup B$, so does its twin term (here the twin of a constant $d$ is $d'$ 
and the twin of $d'$ is $d$).\footnote{
To ensure the latter, we can just add literals like $d'=d'$ to $A$ or  $B$, if needed.} 
 Second, since the number of $\Sigma_A$-ground literals is finite (modulo the identification of a term like $t''$ with $t$), we can assume that if
a $\Sigma_A$-ground 
literal is entailed (modulo $CL$) by $A\cup B$, then it actually occurs in $A$ (and similarly for $B$): the addition of such entailed literals does not in fact compromize our claim. So, let us make the above
assumptions. Notice that (since there is at least one ground \abcommon term), our hypotheses imply that $A\cup B$ is $CL$-consistent, so no pair of contradictory literals can be there.

We can divide the ground terms occurring in $\Sigma_A\cup \Sigma_B$ into equivalence classes (similarly to what  happens in congruence closure algorithms), according to the equivalence relation that holds among $d_1$ and $d_2$ iff
(i) either they both occur in $A$ and $d_1=d_2\in A$, or (ii) they both occur in $B$ and $d_1=d_2\in B$, or (iii) $d_1$ is \astrict, $d_2$ is \bstrict and there exists an \abcommon $t$ such that $d_1=t\in A$, $d_2=t\in B$, or
(iv) $d_1$ is \bstrict, $d_2$ is \astrict and there exists an \abcommon $t$ such that $d_1=t\in B$, $d_2=t\in A$. Notice that, because of our assumptions, the equivalence class of $a$ is different from the equivalence class of $b$.

Since there are no contradictory literals in $A\cup B$, we can build a $\Sigma_A\cup \Sigma_B$-structure $\cA$ in which all literals from $A\cup B$ are true: the support of $\cA$ is formed by the above equivalence classes, a free constant  is interpreted as the equivalence class it belongs to, the twin $C'$ of an equivalence class $C$  is the equivalence 
class formed by the twin
terms of the terms belonging to $C$; moreover, $C$ is a golden cuff link in $\cA$ iff $G(t)\in A\cup B$ (here $t$ is any term belonging to $C$).
 Notice that $\cA\not \models a=b$. However, we are not done, because $\cA$ may  not be a model of $CL$: the reason is that 
there might be more than one golden pair of cuff links. We now show how to merge all golden pairs of cuff links of $\cA$ and get a model $\cM$ of $CL$ having the required properties, namely such that $\cM\models A, \cM\models B$ and $\cM\not \models a=b$.

Consider two different pairs of golden cuff links $C, C'$ and  $D, D'$ (when we say that they are different as pairs of cuff links, we mean that $C$ is different from both $D$ and $D'$). We claim that if we merge $C$ with $D$ and $C'$ with $D'$ as equivalence classes (i.e. if we identify them  as elements from the support of $\cA$), we still have that the literals from $A$ and $B$ are true. In fact, this could possibly be not the case if there are $t\in C, u\in D$ such that $t\neq u\in A\cup B$.  However, literals in $A\cup B$ are all pure, so that either $t,u\in \Sigma_A$ or $t, u\in \Sigma_B$. 
Suppose  $t,u\in \Sigma_A$ (the other case is symmetric); by the construction of $\cA$ and since $C, D$ are golden, we have that $G(t), G(u)\in A\cup B$ and hence (by~\eqref{eq:unique}) the entailed literal $t=u'$ belongs to $A$, so that $C=D'$ which means that $C, C'$
and $D, D'$ are not different pairs of cuff links.

In conclusion, whenever we pick two different pairs of golden cuff links $C, C'$ and  $D, D'$ from the support of $\cA$, we can merge $C$ with $D$ and $D$ with $D'$, without compromizing the truth of $A\cup B$; notice, however, that we can make the symmetric operation and merge $C$ with $D'$ and $C'$ with $D$, again keeping 
the literals in $A\cup B$ true. In the end, we can merge all golden pairs of cuff links into a single one; if $a$ and $b$ belong to $C$ and $D$, respectively, and if $C, D$ are both golden,  we can choose the appropriate merging among the two possible ones, so that in the end we have that $D$ is equal to $C'$, which implies that $a$ and $b$ remains interpreted as different elements in the support of the final model.  
\end{proof}

From the above results and Theorem~\ref{prop:strong_amalgamation_needed} , we obtain:

\begin{corollary}
 $CL$ has the quantifier-free interpolation property and the YMc property, but $CL\cup \EUF$ does not have the quantifier free interpolation property (if the signature of \EUF has at least a unary predicate symbol).
\end{corollary}

A direct counterexample to the quantifier-free interpolation property for the combined theory $CL\cup \EUF$ can be easily obtained by considering the following mutually unsatisfiable  sets of ground literals
$$
A:=\{ G(a), P(a), P(a')\}, \qquad B:=\{ G(b), \neg P(b), \neg P(b')\}
$$
(here $P$ is the extra free predicate). 
%

\end{document}